\documentclass[12pt]{article}
\usepackage{times}

\usepackage{color} 
\usepackage{booktabs}
\usepackage{lineno}
\usepackage{url}
\usepackage{pdfpages}
\usepackage{setspace} 
\usepackage{lineno}
\usepackage{grffile}
\usepackage{amsmath}
\usepackage{amssymb}
\usepackage[comma,sort&compress]{natbib}
\usepackage{hyperref}
\usepackage{float}
\hypersetup{
    bookmarks=true,         % show bookmarks bar?
    unicode=false,          % non-Latin characters in Acrobatâ????s bookmarks
    pdftoolbar=true,        % show Acrobatâ????s toolbar?
    pdfmenubar=true,        % show Acrobatâ????s menu?
    pdffitwindow=false,     % window fit to page when opened
    pdfstartview={FitH},    % fits the width of the page to the window
    pdftitle={My title},    % title
    pdfauthor={Author},     % author
    pdfsubject={Subject},   % subject of the document
    pdfcreator={Creator},   % creator of the document
    pdfproducer={Producer}, % producer of the document
    pdfkeywords={keywords}, % list of keywords
    pdfnewwindow=true,      % links in new window
    colorlinks=true,       % false: boxed links; true: colored links
    linkcolor=blue,          % color of internal links
    citecolor=blue,        % color of links to bibliography
    filecolor=magenta,      % color of file links
    urlcolor=cyan           % color of external links
}

\usepackage[labelfont=bf,labelsep=period,justification=raggedright]{caption}

\makeatletter
\renewcommand{\@biblabel}[1]{\quad#1.}
\makeatother

\date{}

\title{Promoting Fair Proposers, Fair Responders or Both? Cost-Efficient Interference in the Spatial Ultimatum Game}

\author{Theodor Cimpeanu$^{1}$, Cedric Perret$^{2}$, and The Anh Han$^{1,\star}$}
\begin{document}
	\maketitle
	{\footnotesize
		\noindent
		$^{1}$ School of Computing, Engineering and Digital Technologies,  Teesside University, Middlesbrough, UK TS1 3BA\\
		$^{2}$ College of Life and Environmental Sciences, University of Exeter, Exeter, UK EX4 4PY \\
		$^\star$ Corresponding author: The Anh Han (T.Han@tees.ac.uk) %and Tom Lenaerts (Tom.Lenaerts@ulb.ac.be)
	}

\newpage

\section*{Abstract}

Institutions and investors face the constant challenge of making accurate decisions and predictions regarding how best they should distribute their endowments. The problem of achieving an optimal outcome at minimal cost has been extensively studied and resolved using several heuristics. However, these works usually fail to address how an external party can target different types of fair behaviour or do not take into account how limited information can shape this complex interplay. Here, we consider the well-known Ultimatum game in a spatial setting and propose a hierarchy of interference mechanisms based on the amount of information available to an external decision-maker and desired standards of fairness. Our analysis reveals  that monitoring the population at a macroscopic level requires more strict information gathering in order to obtain an optimal outcome and that local observations can mediate this requirement. Moreover, we identify the conditions which must be met for an individual to be eligible for investment in order to avoid unnecessary spending. We further explore the effects of varying mutation or behavioural exploration rates on the choice of investment strategy and total accumulated costs to the investor. Overall, our analysis provides new insights about efficient heuristics for cost-efficient promotion of fairness in societies. Finally, we discuss the differences between our findings and previous work done on the PD and present our suggestions for promoting fairness as an external decision-maker. 
\\

 \noindent \textbf{Keywords:} Ultimatum game, interference, evolutionary game theory, complex network. % \ty{dynamics of innovation and cooperation}.

 \newpage

%%% The code below was generated by the tool at http://dl.acm.org/ccs.cfm.
%%% Please replace this example with code appropriate for your own paper.

%%% Use this command to specify a few keywords describing your work.
%%% Keywords should be separated by commas.

%%%%%%%%%%%%%%%%%%%%%%%%%%%%%%%%%%%%%%%%%%%%%%%%%%%%%%%%%%%%%%%%%%%%%%%%

%%% Include any author-defined commands here.
         
\newcommand{\BibTeX}{\rm B\kern-.05em{\sc i\kern-.025em b}\kern-.08em\TeX}

%%%%%%%%%%%%%%%%%%%%%%%%%%%%%%%%%%%%%%%%%%%%%%%%%%%%%%%%%%%%%%%%%%%%%%%%

\section{Introduction}
The problem of how collective behaviour, such as cooperation, coordination, safety compliance and fairness among self-interested individuals, emerges in evolving, dynamical systems has fascinated researchers from many disciplines, ranging from Evolutionary Biology, Economics, Physics,  Social Sciences and Computer Science \citep{perc2010coevolutionary,key:Sigmund_selfishnes,airiau2014emergence,HanBook2013,nowak:2006bo,West2007,tuyls2007evolutionary,maynard-smith:1982to,han2019modelling}. 
Several mechanisms that are responsible for promoting the emergence of cooperation have been proposed, including direct reciprocity \citep{key:trivers1971,key:Sigmund_selfishnes}, kin selection \citep{Hamilton1964} and network reciprocity \citep{Ohtsuki2006ANetworks} (for review, see \citep{Nowak2012EvolvingCooperation,West2007}). In these works, the evolution of desired collective behaviour is typically shaped by the combined actions of individuals within the systems. 

On the other hand, external interference, where the advocating of certain desired collective behaviour is carried out by an external decision maker (who  does not belong to the system), studies how this can be done in a cost-effective way \citep{chen2015first,han2018cost,Hanijcai2018,wang2019exploring,cimpeanu2019exogenous}. These works aim to identify a broad class of interference strategies, or heuristics, that exploit available information such as global statistics (population behavioural composition), as well as local information such as local behaviour profile and diversity, and graph structures, for budget saving. This line of research is useful to provide insights into the design of self-organised and distributed Multi-Agent Systems (MAS), in order to ensure agents achieve a desired collective state. For instance, one might consider a hybrid system consisting of humans and intelligent machines, in which it is important to ensure a cooperative and trustful relationship amongst each other  \citep{paiva2018engineering,santos2019evolution,Andras2018TrustingSystems}. Another example is how international agencies such as the European Union and United Nations might advocate certain  preferred political behaviours or resolve international conflicts, given a limited budget (e.g. in terms of cost and military resources) \citep{marton2007peace,smidt2020united}.  

The literature on external interference in  evolving, dynamical systems (or populations)  has so far focused on cooperation dilemmas, namely the Prisoner's Dilemma (PD) \citep{han2018cost,Hanijcai2018,cimpeanu2019exogenous} and the Public Goods Game (PGG) \citep{sasaki2012take,chen2014optimal,chen2015first,wang2019exploring}.
 In these games, the interactions are symmetric and the players' roles are equivalent. 
 However, many real-world and MAS interactions are asymmetric, where players may have different baseline characteristics and/or play different roles in the interactions \citep{tuyls2018symmetric,mcavoy2015asymmetric,bianca2021coordination_agreement}. 
Examples include conflict resolution \citep{selten1978note,smidt2020united}, technology adoption by firms  \citep{bianca2021coordination_agreement}, and  multiparty resource allocations \citep{chevaleyre2005issues,lerat2013evolution}, where participants might have different roles (e.g. proposers/dictators vs responders) or  bargaining power in the decision making process.    
In this asymmetric setting, the external decision maker  might need to take into account the difference among players' underlying characteristics, such as their roles in the interactions, in order to optimise the cost and the level of desired behaviour. 
In particular, we might ask, is it enough to target a subset of the roles to already achieve a sufficiently good outcome, 
since collecting information about all the roles might be (very) costly and time consuming?  
 
 This paper contributes to advancing the state-of-the-art by studying cost-efficient external interference in a spatial Ultimatum Game (UG), a popular bargaining game for investigating fair decision making in many disciplines, such as economics and AI/MAS research,  \citep{guth1982experimental,fehr1999theory,de2008learning,santos2019evolution,de2018social,de2011human,rauwolf2018expectations}. In a standard UG,  players have two different roles, proposer and receiver (or responder), with different bargaining powers (See Methods in Section 3 for a detailed description of the game). We consider the spatial version of the game \citep{page2000spatial} where players are distributed on a network in order to  examine how to exploit the roles' asymmetry in both global and local interference strategies \citep{Hanijcai2018}. 
 
% When considering the problem of fairness promotion, interactions are asymmetric and those involved have different roles. We examine in this paper how interference  will, on the one hand, be influenced by this new information regarding players' role, and on the other hand, exploit it in a cost-efficient way. 

In general, a cost-efficient interference problem  consists of solving a bi-objective optimisation problem \citep{han2018cost,wang2019exploring}, maximising the overall level of desired behaviours (in the long run) while ensuring the total cost spent being within budget and/or minimal. The key challenge is that, for evolving, dynamical systems such as those in the above-mentioned examples,  the system dynamics are shaped by various stochastic and random effects, such as those resulting from behavioural updates and  mutation (behavioural exploration) \citep{traulsen2009exploration,rand2013evolution}. With behavioural updates, such as through social learning or reproduction \citep{nowak:2006bo,key:Sigmund_selfishnes}, undesired behaviours might  resurface over time  whenever interference was not sufficiently strong in the past. Through mutation, these behaviours might do so even when they were extinct.  Hence, the external decision maker needs to take into consideration that they will have to repeatedly interfere in the system, in order to sustain the desired behaviour over time. 
Note however that, for simplicity, previous works have either  omitted mutation \citep{han2018cost,wang2019exploring}, or assumed that it is infinitely small (for analytical treatment) \citep{han2018cost}.  
Mutation (behavioural exploration), where agents can freely experiment with new behaviours, is usually non-negligible in real populations and  has been shown to play an important role in enabling cooperation in the context of social dilemmas \citep{DuongHan2020, antal2009mutation, traulsen2009exploration,key:Hanetall_AAMAS2012,rand2013evolution}.
Thus, the present work will also advance the state-of-the-art in this respect, where we will closely examine how different regimes of  mutation, or agents' propensity for behavioural exploration, influence the manner in which external interference should be carried out. Indeed, our results show that when mutation is sufficiently high, in line with those observed in behavioural experiments, optimal interference strategies can be significantly different. 
%\tb{[TA: I think this is an important point that we should and can easily demonstrate with some over time evolution typical runs! We can compare 1) case with no mutation, 2) small mutation 3) large mutation. It should show that in 1) we stop when homogeneous state is reached, in 2) one might need to continue monitoring and interfere, while 3) one needs to constantly monitor the population, etc ]}

The remainder of the paper is structured as follows. The next section reviews most relevant literature. Section 3 then describes the models and methods  in detail. The paper proceeds with presenting the results and a final discussion. We also include with this paper a Supplementary Information (SI) that includes additional results to support the robustness of the paper findings. 

%This paper contributes to advance this line of research, in the following aspects.  

%Previous works have focus on cooperation dilemmas such as the prisoner's dilemma and the public goods game (CITEs). In these games, the interactions are symmetric and players have a single role. When considering the problem of fairness promotion, interactions are asymmetric and those involved have different roles. We examine in this paper how interference  will, on the one hand, be influenced by this new information regarding players' role, and on the other hand, exploit it in a cost-efficient way. 

%Fairness is important MAS, e.g. multiparty resource allocation, conflict resolution, agent bargaining and negotiations. 

%\section{Related Work}

\section{Models and Methods}

\subsection{Ultimatum Game (UG)}

Agents' interaction is modelled  using the one-shot Ultimatum Game (UG) \citep{nowak2000fairness,page2000spatial}. 
In the UG, two players are offered a chance to win a certain sum of money, normalised to 1,  which  they must divide between each other. One player is elected proposer, and suggests how to split the sum, while the other, the receiver (responder) can accept or reject the deal. If the deal is rejected, neither player receives any part of the initial sum. 
As in \citep{nowak2000fairness,page2000spatial}, we assume that a player is equally likely to perform in one of the roles (proposer or receiver). A player's strategy is defined by a pair of parameters, $p$ and $q$. When  acting  as  proposer,  the player  offers  the  amount $p$, while in a receiver's role, the player rejects any offer smaller than $q$. 

As we focus in this paper on the effect of having multiple roles on interference decision making, we consider a baseline UG model where proposers have two possible strategic offers, a low (L, with  $p=l$) and a high (fair) (H, with  $p = h$) one, where $l < h \in [0,1]$. On the other hand, receivers have two options, a low threshold (L, with $q = l$) and a high threshold (H, with $q = h$). Thus, overall, there are four possible strategies HH, HL, LH and LL (i.e. HL would denote proposing high and accepting any offers, etc.).
Given evidence from several behavioural experiments \citep{guth1982experimental,rand2013evolution}, in which people (almost) never offered more than half of the sum in UG, we assume  $h \leq 0.5$. Particularly, we set $h = 0.5$ and $l = 0.1$, as shown in \citep{page2000spatial}. In this scenario, the strategy LL is evolutionarily stable. We also confirm this result in our simulations, as shown in Figure \ref{fig:baseline_mut} in Supplementary Information (SI) and we note that this result is true for several mutation rates. 

The payoff matrix for the four strategies HH, HL, LH and LL reads (for row player)
{\small
\begin{equation}
    \left(
\begin{array}{cccc}
 1/2 & 1/2 & (1-h)/2 & (1-h)/2 \\
 1/2 & 1/2 &  (1-h+l)/2 & (1-h+l)/2 \\
 h/2 &  (1+h-l)/2 & 0 & (1-l)/2 \\
 h/2 &  (1+h-l)/2 & l/2 & 1/2 \\
\end{array}
\right).
\end{equation}
}
\subsection{Population structure and dynamics}
\label{subsec:models}
We consider a population of agents or individuals on a square lattice of size $Z = L \times L$ with periodic boundary conditions--- a  widely adopted population structure in population dynamics and evolutionary games  \citep{szabo2007evolutionary}. We focus our analysis on the efficiency of various interference strategies in spatial settings, adopting an agent-based model directly comparable with the setup of recent lab experiments on cooperation \citep{rand2014static}. We set $L = 100$ for all our experiments, resulting in a population size $Z = 10^4$.

Initially each agent is designated  as one of the four strategies (i.e. HH, HL, LH, HH), with equal probability. 
At each time step or generation, each agent plays the UG with its (four) immediate neighbours. The score for each agent is the sum of the payoffs in these encounters. At the end of each generation an  agent $A$ with score $f_A$  chooses to copy the strategy of a randomly selected neighbour agent $B$ with score $f_B$ with a probability given by the Fermi rule (i.e. \textit{stochastic update}) \citep{traulsen2006stochastic}: $$(1+e^{(f_A- f_B)/K})^{-1},$$ where $K$ denotes the amplitude of noise in the imitation process \citep{szabo2007evolutionary}. Varying $K$ allows us to capture a wide range of update rules and levels of stochasticity, including those used by humans, as measured in lab experiments \citep{zisisSciRep2015,randUltimatum}. In line with previous works and lab experiments \citep{szabo2007evolutionary,zisisSciRep2015,rand2013evolution}, we set $K = 0.1$ in our simulations. With a given probability $\mu$, this process is replaced instead by a randomly occurring mutation. A mutation is equivalent to behavioural exploration, where the individual makes a stochastic decision switch to one of the four available strategies.  

Although our analysis below will focus on the stochastic update rule (in order to examine how stochasticity affects interference, as discussed above), we will also provide results for  \textit{deterministic update} to have a clear comparison with previous works (see e.g. \citep{Hanijcai2018}). For the deterministic update, an agent's strategy is always changed to that of its highest scoring neighbour \citep{nowak1992evolutionary,szabo2007evolutionary}. This is a way of approximating the stochastic update rule where the stochastic effect  is infinitely small, i.e. $K \rightarrow 0$.

%Our analysis will be primarily based on  this  deterministic, standard evolutionary process  in order to focus on understanding the cost-efficiency of different interference strategies.
%However, we confirm that all our conclusions remain valid for a stochastic update  rule (see Section \ref{subsec:stochasticUpdate} for more details).
%Strategies are transmitted through an imitation mechanism, where successful strategies are imitated more often. The success of player A is measured in terms of its cumulative payoff, P(A). The cumulative payoff is the sum of the payoffs earned in the first-neighbor interactions. An imitation step consists of two parts. First, a focal player, A, and one of its next-nearest neighbours, B, are randomly chosen and their payoffs P(A) and P(B) are calculated. Then, the focal player A imitates the strategy of B with probability

%We also consider the deterministic update where each player copies the strategy of the neighbour with highest payoff. 

%We  consider a square lattice graph with periodic boundary conditions (each player has 4 neighbours), of size $100 \times 100$.  The population size is thus $Z = 10^4$. 

We simulate this evolutionary process until a stationary state or a cyclic pattern  is reached. Similarly to  \citep{nowak1992evolutionary}, all the simulations in this work (described in next sections) converge quickly to such a state. For the sake of a clear and fair comparison, all simulations are run for 500 generations. Moreover, for each simulation, the results are averaged over the final 25 generations, in order to account for the fluctuations characteristic of these stable states. 
Furthermore, to improve accuracy, for each set of parameter values, the final results are obtained from averaging 30 independent realisations. When shown in figures, the error bars represent the standard error of the mean between replicates.

Note that in the special case of deterministic update (where we also do not consider mutations), simulations can stop early when the proportion of fair proposers reaches $100\%$. We note that when maximum fairness is not reached, investment can still be ongoing beyond 500 generations and thus, the total cost of interference is dependent on the chosen stopping point. However, our results show that the average investment at the 500 generation mark is never more than $0.2\%$ of the average total investment, for all types of interference. Thus, this arbitrary number has a limited effect and should not affect these results qualitatively. 

%Note that we do not consider mutations or explorations in this work. Thus, whenever the population reaches a homogeneous state (i.e. when the population consists of 100\% of agents adopting the same strategy), it will  remain in that state regardless of interference. Hence, whenever detecting such a state, no further interference will be made. 
 %It is noteworthy that the results remain robust assuming a sufficiently small mutation rate, allowing the population to reach a stable state before any new mutants arise. %This is a common assumption in EGT \citep{HanJaamas2016,szabo2007evolutionary}. FcS: didn't get exactly what you were mentioning here.

%\tr{[Note: maybe for future work only. We add mutation rate to the simulations; that way, the simulations should not stop when reaching 100\% of a strategy since mutation can happen later on. The mutant might evolve if interference is not there.  ]}

\subsection{Cost-Efficient Interference in Networks}
We aim to study how one can efficiently interfere in a structured population to achieve high levels of fairness while minimising the cost of interference. Naturally, the level of fairness is measured by the fraction of fair offers in the population \citep{rand2013evolution}, which is the total of HH and HL frequencies.
An investment decision  consists of a cost  $\theta > 0$ to the external decision-making agent/investor, this value $\theta$ is added as surplus to the payoff of each suitable candidate.

We examine and compare different approaches of interference to induce fairness, based on ensuring fairness for either role or both, leading to different desirable behaviours to be targeted 
\begin{itemize} 
    \item[\textbf{(i)}] ensure all proposals are fair, thus investing in  HH and HL (\textbf{Target: HH, HL});
    \item[\textbf{(ii)}] ensure only fair offers are accepted, thus investing in HH and LH (\textbf{Target: HH, LH}); 
    \item[\textbf{(iii)}] ensure both (i) and (ii), i.e. investing  in HH only (\textbf{Target: HH}).    
\end{itemize}

Moreover, in line with previous works on network interference \citep{chen2015first,han2018cost,cimpeanu2019exogenous},  we will  compare  global  interference strategies where investments are triggered based on network-wide information, and the local ones where investments are based on local neighbourhood information. 

In the \textit{population-based} approach,  a decision to invest in desirable behaviours is based on the current composition of the population. We denote $x_f$ the  fraction  of individuals  in the population with a desirable behaviour, given a targeting approach at hand, i.e. (i), (ii) or (iii) as defined above. Namely, investment is made if $x_f$ is less or at most equal to a threshold $p_f$ (i.e. $x_f \leq p_f$), for $0 \leq p_f \leq 1$. They do not invest otherwise (i.e. $x_f > p_f$).  The value $p_f$ describes how rare the desirable behaviours should be to trigger external support.  
In the \textit{neighbourhood-based} approach,  a decision to invest is based on the fraction  $x_f$ calculated at local level. Investment happens if the proportion of neighbours of a focal individual with the desirable behaviours is less or at most equal to  a threshold $n_f$ (i.e. $x_f \leq n_f$), for $0 \leq n_f \leq 1$; otherwise, no investment is made.  %By varying $n_C$ we can provide an answer to the important question of how cooperative a neighbourhood needs to be for the investor to save the interference cost (i.e. does not interfere) and under which conditions.  %For instance, one can ask whether  it is safe to withdraw  

\section{Results}

When choosing to invest in a population of individuals in an effort to ensure some form of desirable outcome, an external decision maker must first consider several factors before any decision is made. Among these, we consider and aim to resolve the questions regarding what sort of behaviour they should invest in, how large the individual endowment must be, but also what an investor can do when information about the population or the environment is incomplete, or even unknown. As such, we consider that the simplest form of information gathering evaluates the overall population (in the form of some metric measuring fairness on average), as opposed to fine-grained observations on individual neighbourhoods. Likewise, we consider that ensuring all proposals are fair (i.e. investing in HH or HL) is less demanding on an external decision-maker than ensuring that only fair offers are accepted (i.e. investing in HH and LH), which is, in turn, a simpler endeavour than for both the former and latter to be strictly enforced (choosing to invest in HH only). In this way, we can conceptualise a hierarchy of investment strategies, in terms of complexity, some of which may simply be impossible for an investor to follow, merely due to lack of information, funding, or a combination of the two. 

We consider that there exists a \textit{minimal level of fairness} which the external decision maker is  aiming to enforce in regards to the population's behaviour \citep{han2018cost}, and we study the least expensive investment strategies for differing preferences of such an acceptable fairness.

\begin{figure}
    \centering
    \includegraphics[width=0.8\linewidth]{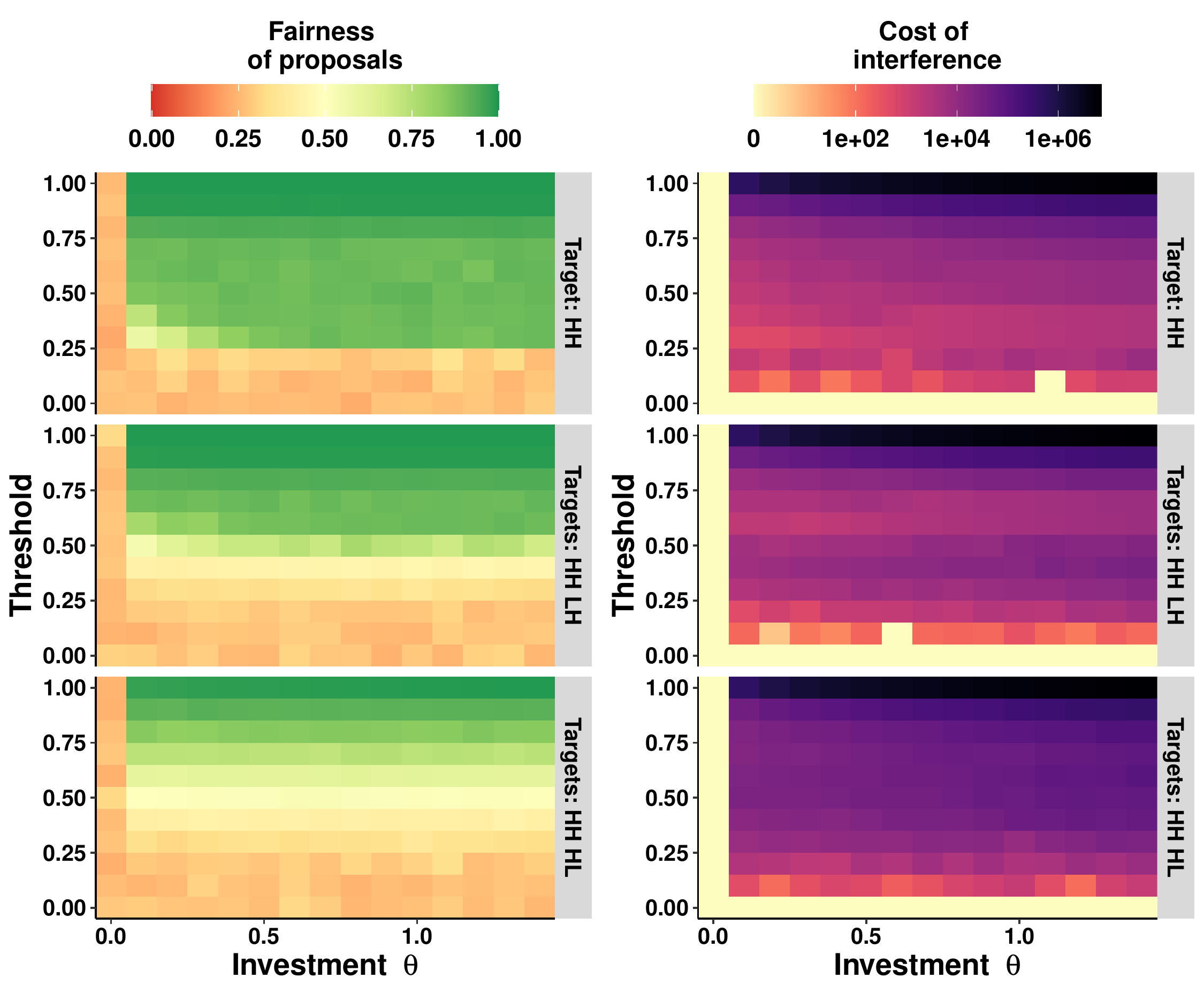}
    \caption{Average fairness (left) and average cost of interference (right) as a function of the individual endowment $\theta$ and the threshold $p_f$ (population-based, $\mu = 0.01$, stochastic update). Each row represents a different targeting scheme. The cost of interference is shown on a logarithmic scale.} 
    \label{fig:popbased_freq_cost_mut0.01}
\end{figure}

\subsection{Population-based results}
Firstly, we explore the simplest class of investment strategies,  using a macroscopic metric of the population, measuring average fairness in the whole system (population). In Figure \ref{fig:popbased_freq_cost_mut0.01}, we clearly observe the difference between the three targets for investment. We would like to point out the higher levels of fairness obtained using the HH targeting scheme, especially for a lower threshold $p_f$. We also notice an increase in the threshold for investment $p_f$ in order to achieve similar levels of fairness. When it comes to the accumulated cost of interference, we see that HH is the most cost-effective solution, due to the previously perceived lower threshold required to maintain fairness. 
\begin{figure}
    \centering
    \includegraphics[width=0.9\linewidth]{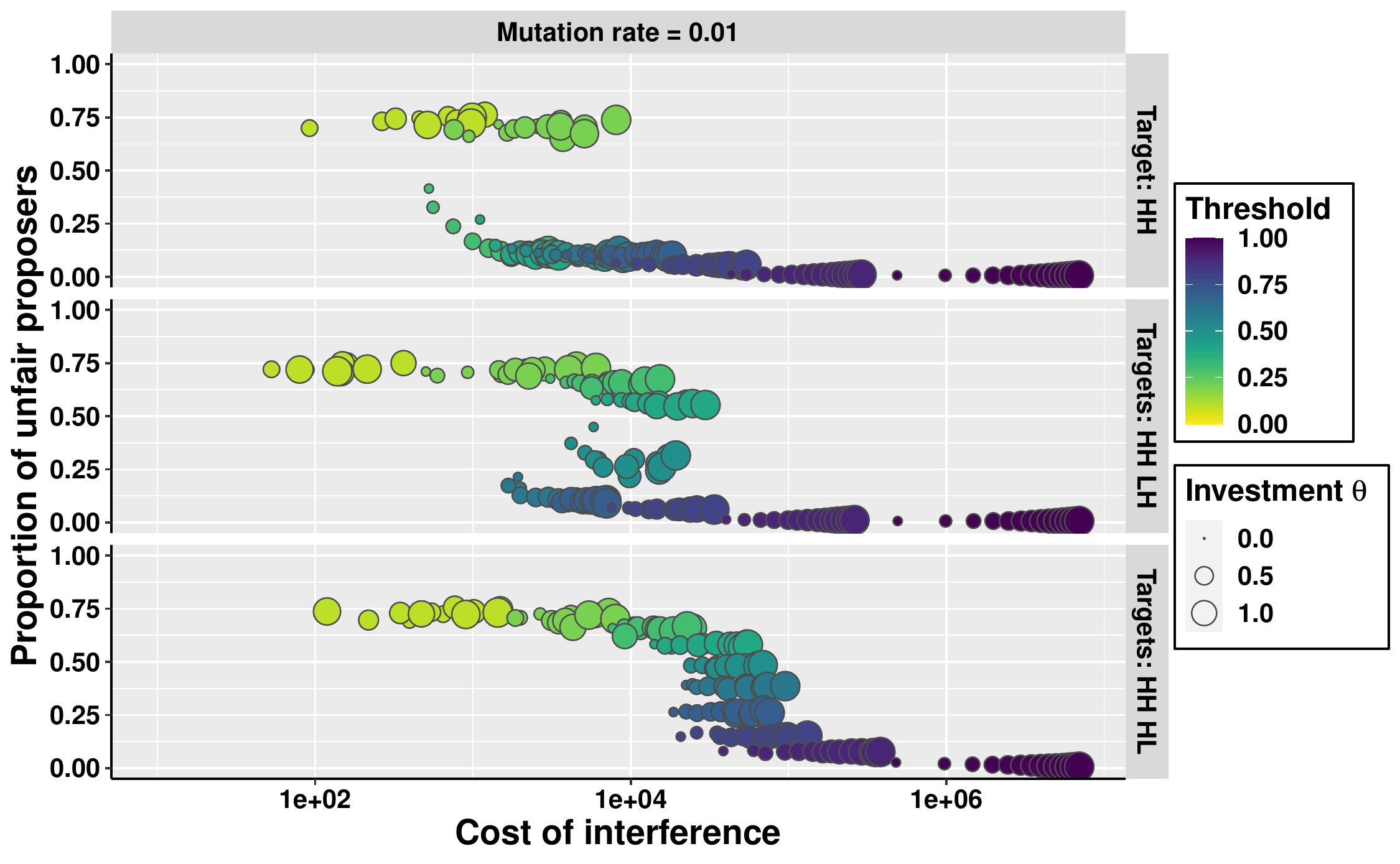}
    \caption{Proportion of unfair proposers as a function of average cost of interference for different targeting scheme  (population-based, $\mu = 0.01$, stochastic update). The size and colour of the circles correspond to investment amount and threshold of investment, respectively. We note that the most desirable outcomes are closest to the origin.}
    %I put unfair so we have a classic pareto front where the desirable outcome is on the bottom left.
    \label{fig:popbased_pareto}
\end{figure}
\begin{center}	
\begin{table}[t]
\centering
\caption{Most cost-efficient  scheme to reach a minimum fairness of proposals for different mutation rates (population-based, stochastic update). There exists no schemes which satisfy the higher minimum fairness requirements in the case of very high mutation rate, written as `--' in the table.}
\label{table:pop-based}
\small
\begin{tabular}{c c c c c c} \toprule
Mutation  rate & Minimum  fairness & Target & Threshold & $\theta$ & Cost\\ \midrule
 $10^{-4}$ & 75\% & HH & 0.3 & 0.1 & 530\\  
 $10^{-4}$ & 90\% & HH & 0.3 & 0.1 & 530\\  
 $10^{-4}$ & 99\% & HH & 0.3 & 0.4 & 999\\  
 $10^{-2}$ & 75\% & HH & 0.3 & 0.3 & 750\\  
 $10^{-2}$ & 90\% & HH & 0.3 & 0.7 & 1747\\  
 $10^{-2}$ & 99\% & HH & 1 & 0.1 & 487514\\
 0.2 & 75\% & HH & 0.6 & 0.2 & 358089\\  
 0.2 & 90\% & -- & -- & -- & --\\ 
 0.2 & 99\% & -- & -- & -- & -- \\ \bottomrule
\end{tabular}
%I remove 25% because we reach it without interference
\end{table}
\end{center}
Figure \ref{fig:popbased_pareto} further exemplifies the finding that targeting HH is the optimal scheme for population-based interference. Each row (portraying the different targeting schemes), drifts further away from the cost-optimal bottom left. As the threshold increases, so does the total cost, so the regions of high fairness for a lower threshold observed in Figure \ref{fig:popbased_freq_cost_mut0.01} coincide with the maximal savings (while still achieving desired levels of fairness).

Table \ref{table:pop-based} shows the most cost-efficient schemes for ensuring specific standards of fairness when only a population-based approach is possible, under differing rates of mutation ($\mu$). We observe  a definitive bias towards the most complex investment scheme (i.e. targeting HH players), which reiterates our previous observation. We note that, in order to maintain a desired level of fairness, an external decision maker must increase the threshold at which they resume their investment, but also the individual endowment ($\theta$). It becomes increasingly difficult to maintain standards of fairness when the population is exposed to high degrees of behavioural exploration and this naturally attracts an increase in total cost for the investor. We report similar figures for other values of $\mu$ in Figures \ref{fig:pop_freq_cost_different_muts}, \ref{fig:pop_freq_different_muts}, \ref{fig:pop_pareto_different_muts}, in SI.

Moreover, we observe  an increase in fairness for all schemes of interference, across most values of individual endowment $\theta$, which bodes well when the external decision maker possesses limited knowledge. If reducing cost is not the main objective, fairness can be maintained using any targeting scheme (i.e. any relevant observations made about the population), by increasing the minimum threshold $p_f$. 

When the external decision maker is limited to the macroscopic metrics associated with population-based interference, interference is characterised by its strictness. To elaborate, information gathering should be the main goal for the investor, as ensuring that proposals and responses are simultaneously fair (i.e. targeting HH) is the optimal outcome. In this way, the minimum threshold can be kept low, reducing the accumulated cost.

\begin{figure}
    \centering
    \includegraphics[width=0.8\linewidth]{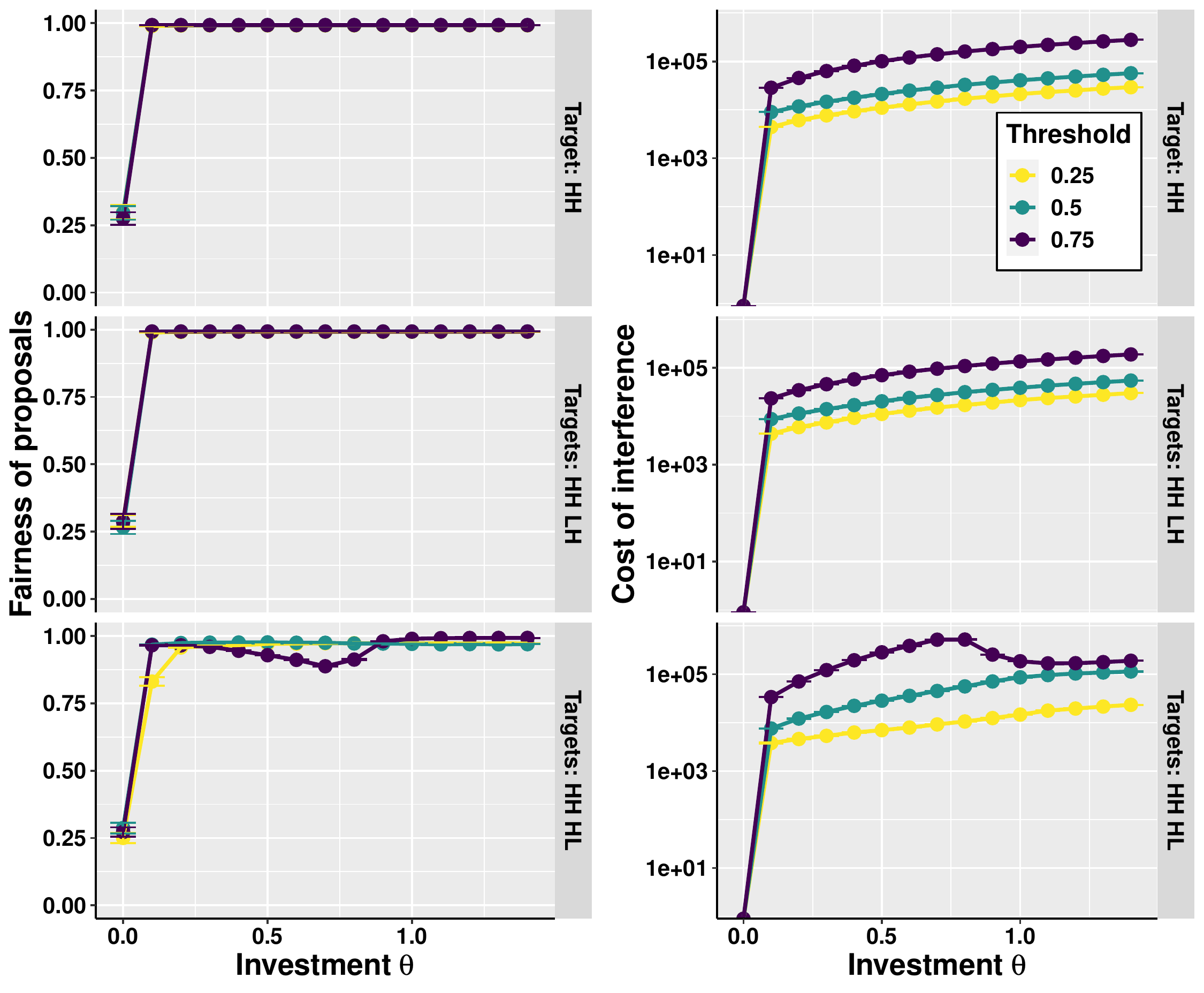}
    \caption{Average fairness  (left) and  average cost of interference (right) as a function of $\theta$ and  threshold $n_f$   (neighbourhood-based, $\mu = 0.01$, stochastic update). Each row represents a different targeting scheme. The cost of interference is on a logarithmic scale.}
    \label{fig:nebbased_freq_cost_mut0.01}
\end{figure}

%PARETO PLOT: Put the x-axis between 0 and max?

%\tr{[TA: maybe we can discuss this result in relation to the PD results, e.g. in the ijcai 2018 paper (for lattice): much stricter investment schemes are required for PD (best threshold is close to 100\%). That work did not consider mutation, while we shows it is an important factor here (We should emphasise this point as one of  key novelties in this paper).     ]}

\subsection{Neighbourhood-based interference}

%I need the sd of freq of HH and HL to put error bar

Previous works on the PD have shown that the greatest gains in cooperation (while maintaining a minimal investment cost) require very detailed observations of individual neighbourhoods, coupled with overly strict investment schemes \citep{han2018cost,cimpeanu2019exogenous,Hanijcai2018}.
%very detailed observations of individual neighbourhoods, coupled with a very strict investment scheme can produce the greatest gains in cooperation with minimal cost required to be invested, in the Prisoner's Dilemma . 
In order to decipher whether or not these findings hold for the spatial ultimatum game, we study the outcome when an investor can perceive fairness at the local level. 

Figure \ref{fig:nebbased_freq_cost_mut0.01} reports the relationship between gains in fairness and increases in cost for an external investor, with diverse targets for receiving investments. We  observe that fairness is more easily achieved than in population based interference, with only a very low investment required to sustain a majority of fair proposals. Further investment increases the cost of interference, but only slightly. If different thresholds result in fairness, Figure \ref{fig:nebbased_freq_cost_mut0.01} shows that a threshold of $25\%$ is the most cost-efficient. Similarly to population-based interference, the external decision maker should invest only when a large proportion of unfair individuals are present to limit the cost of investment. Finally, there are no significant differences between targeting schemes. 

Similarly to our findings using a population-based approach, we observe  that the more prohibitive option, HH, is also the most cost-effective. On the other hand, high fairness can be achieved in all three cases for the same values of endowment. Ensuring that all proposals are fair (thus investing in HH and HL), can lead to an increase in cost of interference, and a decrease in fairness gains (relative to the other two interference strategies). While all investment schemes evidently succeed in promoting the evolution of fairness, only ensuring the equitable proposals is not as reliable as encouraging discerning responses to offers or both. We note that this effect can only be seen when the threshold for investment is very high (i.e. an investor only invests in neighbourhoods with three or more fair proposers). As discussed earlier, investing in neighbourhoods with  at most one fair agent and not investing otherwise,  solves this dilemma. 

\begin{figure}
    \centering
    \includegraphics[width=0.8\linewidth]{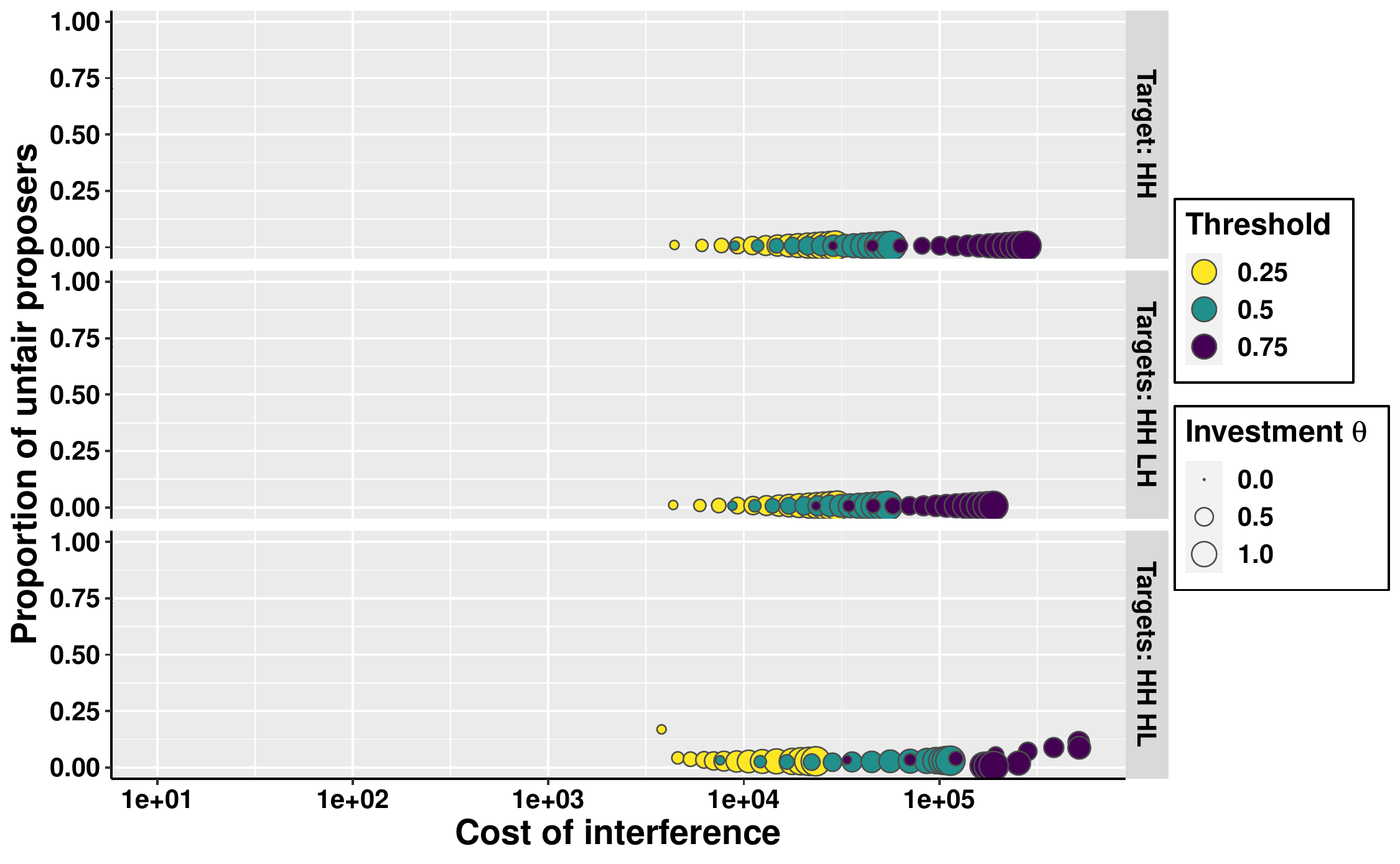}
    \caption{Proportion of unfair proposers as a function of average cost of interference for different targeting scheme  (neighbourhood-based, $\mu = 0.01$, stochastic update). The size and colour of the circles correspond to investment amount and threshold of investment, respectively. We note that the most desirable outcomes are closest to the origin.}
    \label{fig:nebbased_pareto}
\end{figure}

Markedly, it is not effective to invest in neighbourhoods with a high percentage of fair proposals. These results point to a key observation, that it is more important to invest in fair proposers when there are few of them in a specific neighbourhood. In this sense, the lonely fair individuals require aid in otherwise competitive, unjust entourages. This result can further be seen in Figure  \ref{fig:nebbased_pareto}. By being very selective with which neighbourhoods the external investor chooses to invest in (i.e. only choosing very fair neighbourhoods), they inadvertently produce a much higher final cost to their own selves. An external decision-maker would then unwittingly keep investing in fair proposals ad infinitum because fairness is eventually reached in the ultimatum game, even when individual endowment is relatively low. It is clear, therefore, that to reduce potential costs, only players in unfair groups should be eligible for investment. Therefore, the defining characteristic of neighbourhood-based interference is the low threshold for investment ($25\%$).
\begin{center}
\begin{table}[t]
\centering
 \caption{Most cost-efficient  scheme  to reach a minimum fairness of proposals for different mutation rates (neighbourhood-based, stochastic update). There exists no schemes which satisfy the higher minimum fairness requirements in the case of very high mutation rates, written as `--' in the table.} 
 \label{table:neb-based}
\small
\begin{tabular}{c c c c c c}\toprule
Mutation rate & Minimum  fairness & Target & Threshold & $\theta$ & Cost\\  \midrule
 $10^{-4}$ & 75\% & HH & 0.25 & 0.1 & 1395\\  
 $10^{-4}$ & 90\% & HH & 0.25 & 0.1 & 1395\\  
 $10^{-4}$ & 99\% & HH & 0.25 & 0.1 & 1395\\  
 $10^{-2}$ & 75\% & HH HL & 0.25 & 0.1 & 3794\\  
 $10^{-2}$ & 90\% & HH LH & 0.25 & 0.1 & 4352\\  
 $10^{-2}$ & 99\% & HH LH & 0.25 & 0.2 & 5957\\
 0.2 & 75\% & HH & 0.25 & 0.4 & 150777\\  
 0.2 & 90\% & -- & -- & -- & --\\ 
 0.2 & 99\% & -- & -- & -- & -- \\ \bottomrule
\end{tabular}
\end{table}
\end{center}
By varying minimal fairness requirements and rates of mutation, we can gain further insight into which investment strategies are the most robust and cost-effective. Table \ref{table:neb-based} highlights some surprising findings. We see that neighbourhood based interference can result in a higher total cost than the optimal population-based interference schemes (see Table \ref{table:pop-based}). Previous work has shown that more specific and restrictive intervention schemes are more effective in the PD \citep{Hanijcai2018,cimpeanu2019exogenous}, but by being able to target different roles in the Ultimatum game, these differences can be mitigated. Furthermore, mutation rate serves as an equaliser between the investment targets and we observe that less specific schemes (HH \& HL and HH \& LH) are the most cost-efficient options. We note that the differences between results are small enough that different runs could yield any outcome in the case of high or intermediate mutation rates. The lack of significant variability among the distinct targeting schemes contrasts strongly with the findings on the PD \citep{han2018cost,cimpeanu2019exogenous}. We report similar figures for other values of $\mu$ in Figures \ref{fig:neb_freq_different_muts}, \ref{fig:neb_freq_cost_different_muts}, \ref{fig:neb_pareto_different_muts}, in SI.

\begin{figure}
    \centering
    \includegraphics[width=0.8\linewidth]{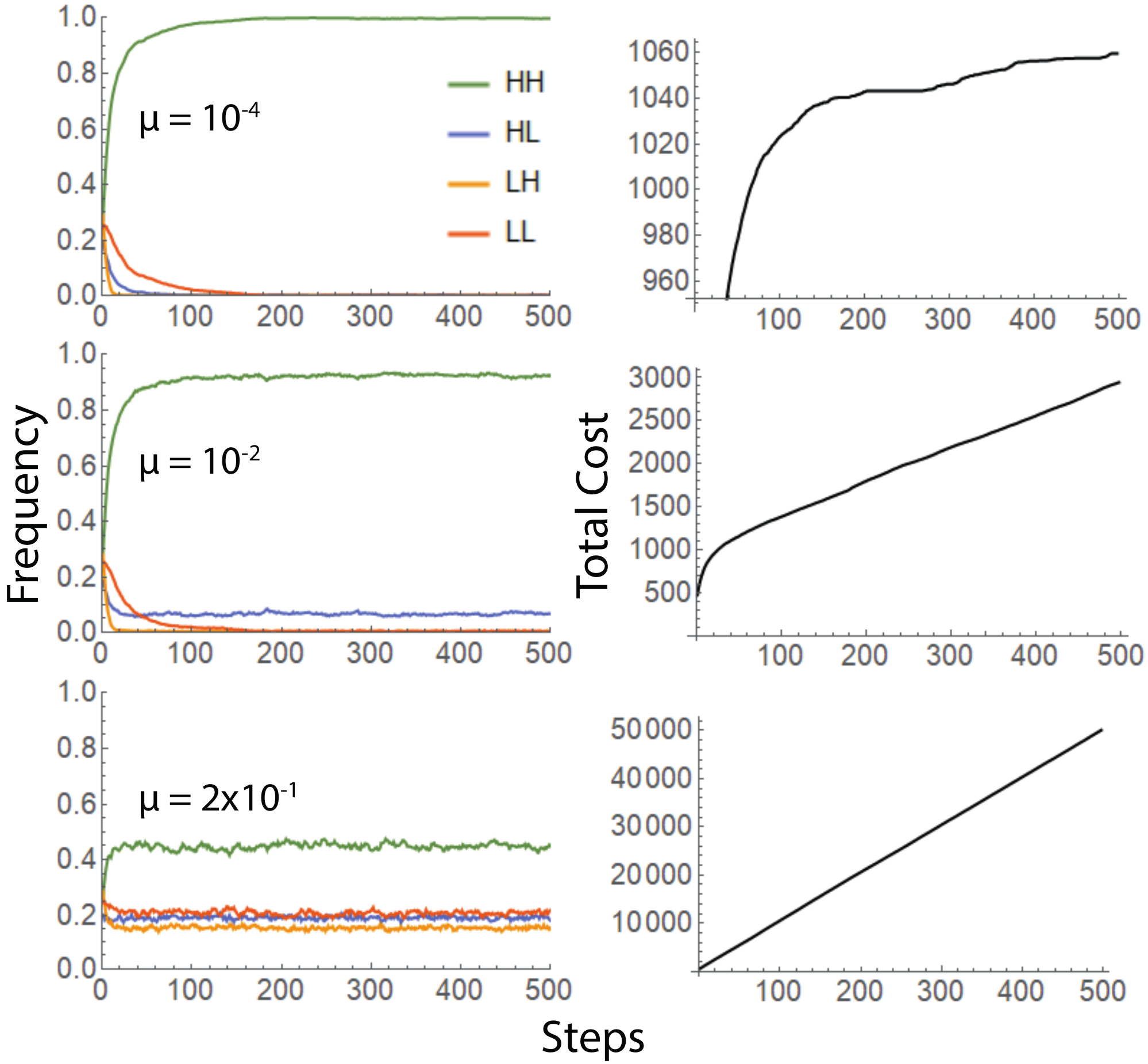}
    \caption{Typical runs showing the evolution of fairness and the associated total cost of interference for various mutation rates ($\mu$) (neighbourhood-based, stochastic update). Parameters:  $\ n_f = 0.25, \ \theta = 0.1, \ Target = HH.$ The choice of parameter values was motivated by selecting the optimal solutions in Table \ref{table:neb-based}.}
    \label{fig:time-evolution-lowthresh}
\end{figure}

\begin{figure}
    \centering
    \includegraphics[width=0.8\linewidth]{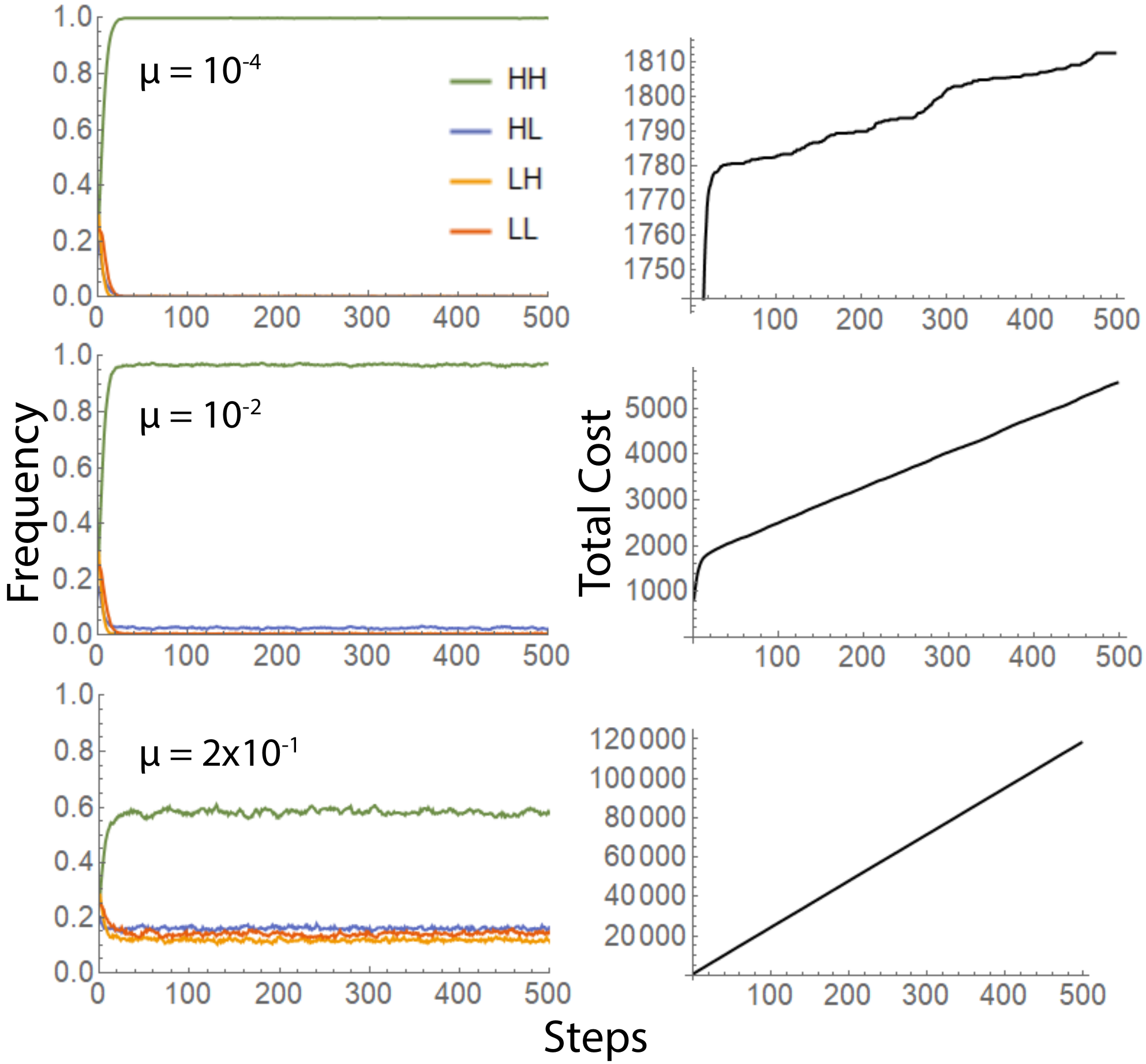}
    \caption{Typical runs showing the evolution of fairness and the associated total cost of interference for various mutation rates ($\mu$)  (neighbourhood-based, stochastic update). Parameters:  $\ n_f = 0.5, \ \theta = 0.1, \ Target = HH.$ The choice of parameter values was motivated by selecting the optimal solutions in Table \ref{table:neb-based}.}
    \label{fig:time-evolution-highthresh}
\end{figure}

\subsection{Evolution of strategies over time}

We make use of the optimal parameter values identified in Tables \ref{table:pop-based}  and \ref{table:neb-based} to explore the evolution of fairness over time for all the strategies in the population, as well as any associated accumulated costs. Through this analysis, we clarify some of the dynamics differentiating the different decisions for investment, as well as the effects of varying mutation rates upon the outcomes and the options available to investors.

The effects of mutation on the optimality of different interference schemes can be seen in Figure \ref{fig:time-evolution-lowthresh}. As the mutation rate ($\mu$) increases, the capacity of maintaining a threshold of fairness decreases (as also seen in Table \ref{table:neb-based}). An external investor must increase their individual investment amount in order to meet these new demands set out by the increased mutation rates, and by doing so they can maintain fairness levels to a respectable standard. 

\begin{figure}
    \centering
    \includegraphics[width=0.8\linewidth]{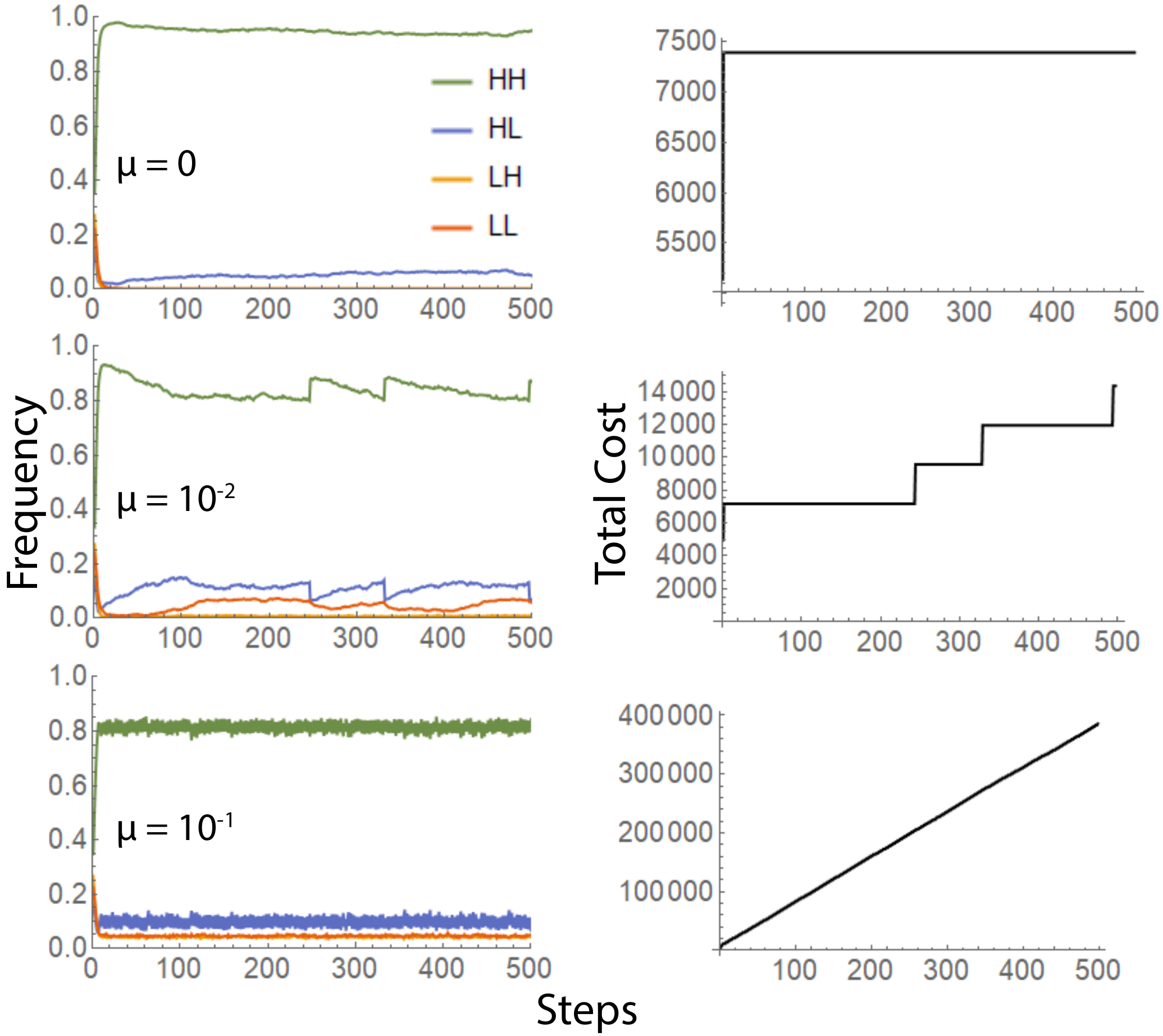}
    \caption{Typical runs showing the evolution of fairness and the associated total cost of interference for various mutation rates ($\mu$)  (population-based, stochastic update). Higher mutation rates leads to  an increasing need for interference  over time. Parameters:  $\ p_f = 0.8, \ \theta = 0.3, \ Target = HH.$ The choice of parameter values was motivated by selecting the optimal solutions in Table \ref{table:pop-based}.}
    \label{fig:time-evolution-mutation}
\end{figure}

To better highlight the sharp increases in the cost associated with the non-optimal  threshold (i.e. when it is greater than $25\%$) for neighbourhood-based interference, we show such typical runs for varying mutation rates for the $50\%$ threshold in Figure \ref{fig:time-evolution-highthresh}. When comparing Figures \ref{fig:time-evolution-lowthresh} and \ref{fig:time-evolution-highthresh}, we note the relative differences in total accumulated costs attributed to the choice of the threshold for investment $n_f$. We also note that increasing rates of behavioural exploration (mutation) amplifies this discrepancy.

\begin{figure}
    \centering
    \includegraphics[width=0.8\linewidth]{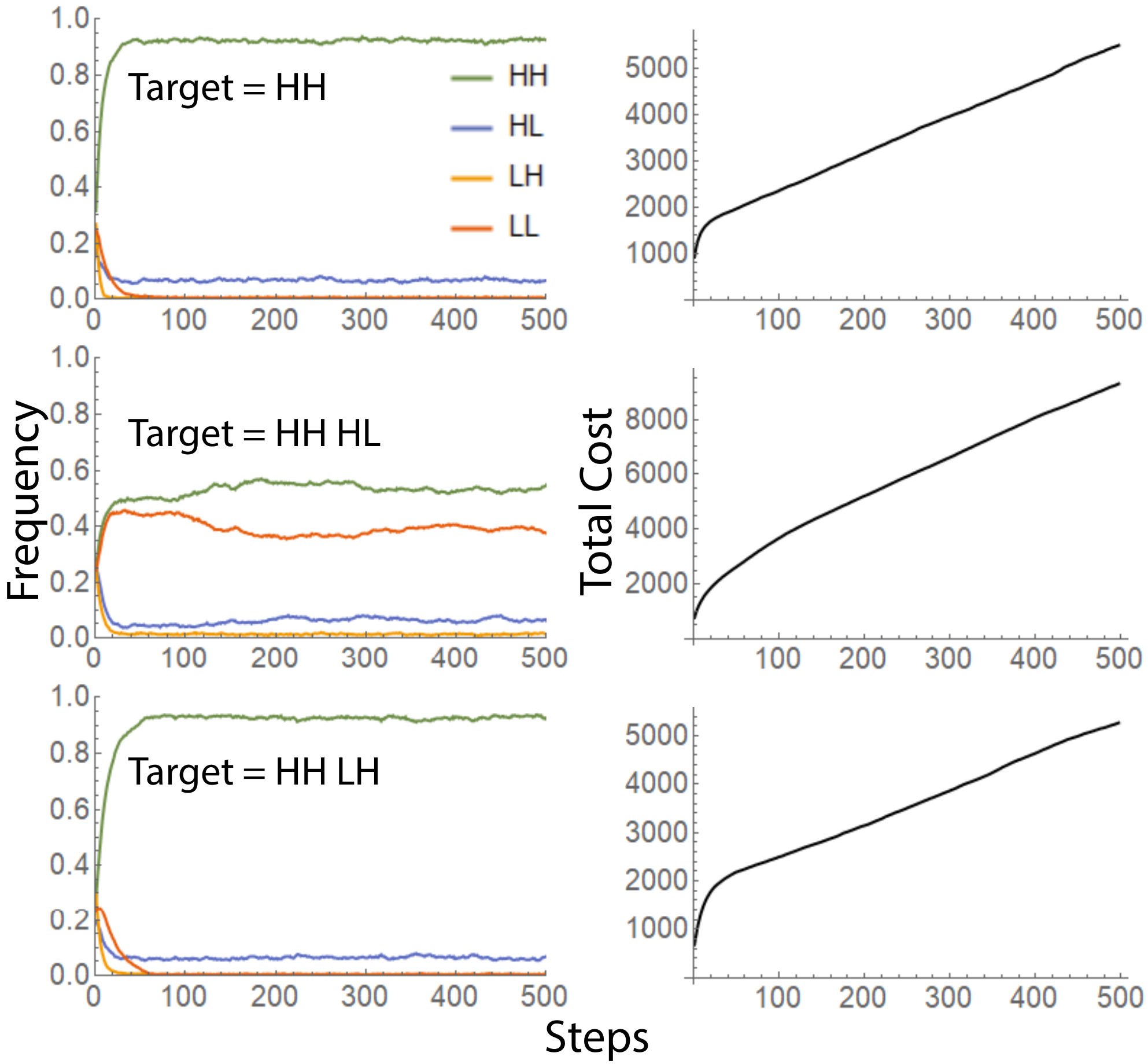}
    \caption{Typical runs showing the evolution of fairness and the associated total cost of interference for various targeting schemes (neighbourhood-based, stochastic update). Parameters:  $\ n_f = 0.25, \ \theta = 0.2, \ \mu = 10^{-2}.$ The choice of parameter values was motivated by selecting the optimal solutions in Table \ref{table:neb-based}.}
    \label{fig:time-evolution-targets}
\end{figure}

We show how less specific interference strategies, which require less information gathering, can be effective in facilitating the evolution of fairness, when local monitoring is possible (Figure \ref{fig:time-evolution-targets}). Promoting fair proposals may often not be sufficient for low individual investment budgets (which are also the optimal solution) --- in such cases fairness does not evolve. This occurs due to the inability of indiscriminate fair proposers to protect themselves against unfair proposers. Investing in fair proposers, in this case, artificially protects them against very competitive selection pressures. 

Figure \ref{fig:time-evolution-mutation} showcases how different mutation rates call for different approaches to interference. 
%For the sake of presentation, we chose a straightforward example --- the first row shows no mutation and in this case we see the system quickly converge to fairness after a brief initial period of investment. After this momentary assistance by the external agent, no more investment is required. In the second row, we show a typical run for intermediate mutation rates - in this case, the investor has to keep coming back and invest when appropriate, this naturally attracts a higher total cost. Finally, in the case of a high mutation rate, we see that the investor must continuously invest and monitor the population, while the total cost skyrockets. 
 As shown previously, optimal interference strategies vary according to the mutation rate. 
 We point out the three different cases in which an investor might find themselves in. First, when few initial rounds of investment are enough for the system to converge and stabilise to a desired state. Second, an investor might be required to reinvest when the population tends to revert back to its initial condition. Lastly, constant investment is required to maintain a desired level of fairness, with the total cost skyrocketing accordingly.
 To some extent, a fair population can better deal with unfair invaders and this explains the need for a sufficiently high initial investment when mutation rates increase.
 
 Finally, behavioural exploration motivates the manner or strength (in terms of individual endowment) of any initial efforts to moderate unfair behaviour. Figures \ref{fig:time-evolution-lowthresh}, \ref{fig:time-evolution-highthresh} and \ref{fig:time-evolution-mutation} show that the increase in cost is linear and ever-growing for high mutation-rates and gradually sharper at the beginning for lower mutation, eventually plateauing when the population is exposed to little or no behavioural exploration. 

\begin{center}
\begin{table}[t]
\centering
 \caption{Most cost-efficient population-based scheme (deterministic update) to reach a minimum fairness of proposals.} 
 \label{table:DETERMINISTIC-POP-based}
\small
\begin{tabular}{c c c c c}\toprule
Minimum  fairness & Target & Threshold & $\theta$ & Cost\\  \midrule
75\% & HH & 0.5 & 0.5 & 1251\\  
90\% & HH & 0.6 & 0.9 & 2228\\  
99\% & HH & 0.9 & 1.1 & 5488\\  
\end{tabular}
\end{table}
\end{center}

\begin{center}
\begin{table}[t]
\centering
 \caption{Most cost-efficient neighbourhood-based scheme (deterministic update) to reach a minimum fairness of proposals.} 
 \label{table:DETERMINISTIC-neb-based}
\small
\begin{tabular}{c c c c c}\toprule
Minimum  fairness & Target & Threshold & $\theta$ & Cost\\  \midrule
75\% & HH & 0.25 & 0.8 & 2146\\  
90\% & HH & 0.25 & 0.8 & 2146\\  
99\% & HH & 0.25 & 1 & 2513\\  
\end{tabular}
\end{table}
\end{center}

\begin{figure}
    \centering
    \includegraphics[width=0.8\linewidth]{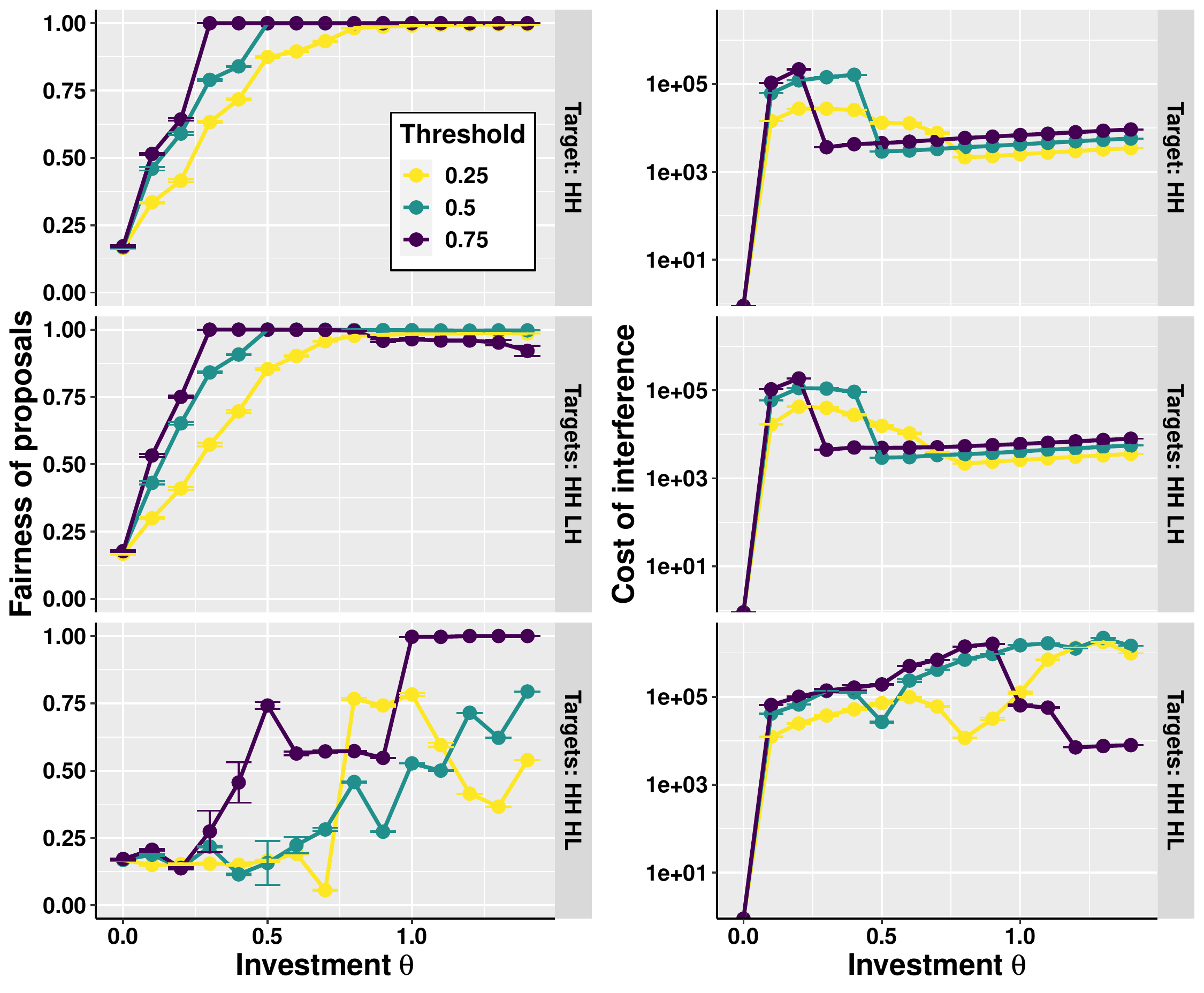}
    \caption{Average fairness (left) and  average cost of interference (right) as a function of $\theta$ and  threshold $n_f$  (neighbourhood-based, deterministic update). Each row represents a different targeting scheme. The cost of interference is on a logarithmic scale for clarity.}
    \label{fig:nebbased_freq_cost_deterministicMain}
\end{figure}

\subsection{Deterministic update}
The results and findings reported so far were based on the stochastic update rule. We now take a step back and consider whether our findings would still hold for the deterministic rule (see again Methods section). It is not only for the sake of a direct comparison with a previous analysis reported in  \citep{Hanijcai2018}, where cost-efficient interference was studied for the spatial PD in a deterministic setting (with no mutation). It would also allow us to examine if the findings above would remain robust for the deterministic update, a popular  approximation for rare stochastic effect (or infinite intensity of selection) that is regularly used in the literature \citep{Szabo2007}. Tables \ref{table:DETERMINISTIC-POP-based} and \ref{table:DETERMINISTIC-neb-based} report results for the optimal interference strategies, in population-based (for a full report, see Figure \ref{fig:popbased_freq_cost_deterministic}, in SI) and neighbourhood-based schemes, respectively. We observe that for both schemes, targeting HH is always the best option. This is the same as the stochastic approach for population-based schemes but different from the neighbourhood-based ones. 
However, for the latter ones, the optimal threshold $n_f = 0.25$ remains the same as in the case of stochastic update, see also Figure \ref{fig:nebbased_freq_cost_deterministicMain}. This is in stark contrast with the PD results where $n_f = 0.75$ was always the optimal choice.

\section{Discussion}
In summary, this paper has advanced the state of the art of the literature on external interference in dynamical systems, or populations of self-interested individuals, in two main respects: i) we have addressed an asymmetric interaction setting, in the form of the Ultimatum game, where players have different roles in the interaction. We have shown that it is crucial to consider the roles' asymmetry to provide cost-efficient investment strategies. This important  analysis  was not possible in previous works where symmetric games were studied \citep{chen2014optimal,han2018cost,Hanijcai2018,chen2015first,wang2019exploring,cimpeanu2019exogenous}; ii) we have incorporated realistic levels of mutation or behavioural exploration in our analysis and have shown that they strongly affect the manner in which interference should be carried out. Previous works have always omitted  mutation or assumed that it is infinitely small, thereby being unable to address this important issue for real-world populations and applications.

We  have identified several  key features that are required for a cost-effective interference scheme. On the one hand, population-based schemes are characterised by the need of extensive information gathering about both roles, as targeting HH always leads to the optimal strategy. On the other hand, neighbourhood-based schemes are characterised by their flexibility, where the optimal strategy always entails that investment is only made when there is at most one player with the desirable behaviour in the neighbourhood (i.e. no investment should be made when there is a half or larger fraction of such behaviour the neighbourhood). Our findings stand out in stark contrast with previous works on cooperation dilemmas, where both population and neighbourhood-based schemes require a highly strict investment approach.

% As mentioned above, the problem of external intervention in evolving, dynamical MAS has been formalised and studied in the context of cooperation dilemmas such as the prisoner's dilemma \citep{han2018cost,cimpeanu2019exogenous,Hanijcai2018} and the public goods game  \citep{chen2015first,wang2019exploring}. In these games, the interactions are symmetric and players have a single role. When considering asymmetric interactions such as the popular ultimatum game, players involved in the interactions have distinct roles. We show that more efficient mechanisms can be devised by taking into account this new information regarding players' role.   

The ultimatum game has been widely studied, whether with theoretical models (see \citep{Debove2016ModelsClassification} for a review) or behavioural experiments (see \citep{Guth2014MoreLiterature} for a review).
The main motivation arises from the gap between theoretical predictions, in which rational individuals keep most of the endowment and the responders accept any positive proposition however small it may be, and experiments, in which individuals propose 40\% to 50\% of their endowment (and often get punished if they propose less) \citep{guth1982experimental}. That said, previous works have investigated how fairness can evolve in models of the ultimatum game, wherein  several mechanisms promoting the emergence of fairness have been identified. We note that we align our definition of fairness with these previous works, where generous proposers are deemed as fair, regardless of their behaviour when acting in the role of the responder.

Among others, Nowak et al. have studied the evolution of fairness in the  Ultimatum game under indirect reciprocity \citep{nowak2000fairness}, i.e. when players can observe others' interactions and have information about the reputations of others. We do not rely on reputation building mechanisms, as the role of this mechanism can be limited in large groups, where one-shot interactions between strangers are common. Page et al. and Sinatra et al.
 have developed spatial models of the Ultimatum game, where interactions happen only between neighbours \citep{page2000spatial,sinatra2009ultimatum}. It has been shown that a spatial structure can promote the emergence of higher levels of fairness, but an equal split between proposers and responders is yet to be reached. The  model developed in the present work  has also considered a spatial model because (i) it captures an essential feature of many  real-world networks of contacts \citep{barabasi2014linked}, and (ii) it allows us to explore the effects of interference localised in particular neighbourhoods, which has been shown to be more cost efficient \citep{Hanijcai2018}. Furthermore, Rand et al. have shown that even if the population is well-mixed, fairness can result directly from the effect of randomness due to mutation and stochastic strategy updates \citep{randUltimatum}. Intuitively, the uncertainty in the responders' choice forces the proposers to offer a high proposal so as to avoid rejection. We show that these stochastic factors  also strongly influence the manner in which external interference may be performed while maintaining cost-efficiency.
Moreover, the Ultimatum game has also been used to study how fairness can emerge  in a hybrid population of human-agent interactions  \citep{de2018social,santos2019evolution}. These works, however, do not consider external interference.

The problem of how to externally influence a system of multiple interacting agents to achieve a certain desired behaviour has been of significant interest in mechanism design, network theory and  control theory literature. 
For example, how to maximise influence  in networks has been studied in  \citep{1500416,bloembergen2014influencing,riehl2016towards}. Moreover, Endriss et al. have investigated how to tax games so as incentivise certain behaviours at system equilibrium \citep{endriss2011incentive}; while,  Wooldridge has presented potential ways to manipulate games in order to achieve desired behaviours \citep{Wooldridge12}. 
These works, however, assume that the decision-maker possesses full control of the agents within the systems. With our approach, the decision-maker has little or no direct control on the agents' behaviour, so we can rely only on rewarding schemes and their effects as ways of motivating the evolution of fairness. It is noteworthy that these works do not focus on the cost efficiency problem, whereas cost optimisation is one of our main goals. 

% \tr{Finally,  % "the cost of stability" 
% Bachrach et al. explored how much investment, in a cooperative game, is needed to ensure that a certain coalition structure is stable \citep{Bachrach2009}.
% This work, however, only considers non-evolving systems with a single time step, and thus, interference can only be applied once.
% In contrast, in our system, due to its evolutionary dynamics, interference is repeatedly carried out over time.}

Our future work will examine other asymmetric games with multiple roles, such as the trust  and anticipation games, where the bargaining nature is different from the ultimatum game \citep{Gut2009,perret2020or,zisisSciRep2015,rauwolf2018expectations}, to see how this bargaining factor might affect the way interference needs to be made. We are also interested in how different network structures influence the interference strategies in asymmetric interactions, which has been studied for symmetric games \citep{cimpeanu2019exogenous}. 

\section*{Acknowledgements}
T.C., C.P. and T.A.H. were supported by Future of Life Institute grant RFP2-154. T.A.H. is also supported by a Leverhulme Research Fellowship (RF-2020-603/9).

\bibliographystyle{apalike}
\bibliography{references}  

\newpage
\renewcommand{\thefigure}{S\arabic{figure}}
 \renewcommand{\thetable}{S\arabic{table}}
 \setcounter{figure}{0}   

\section{Supplementary Information}
\label{section:supplementary}

\begin{figure}[H]
    \centering
    \includegraphics[width=0.9\linewidth]{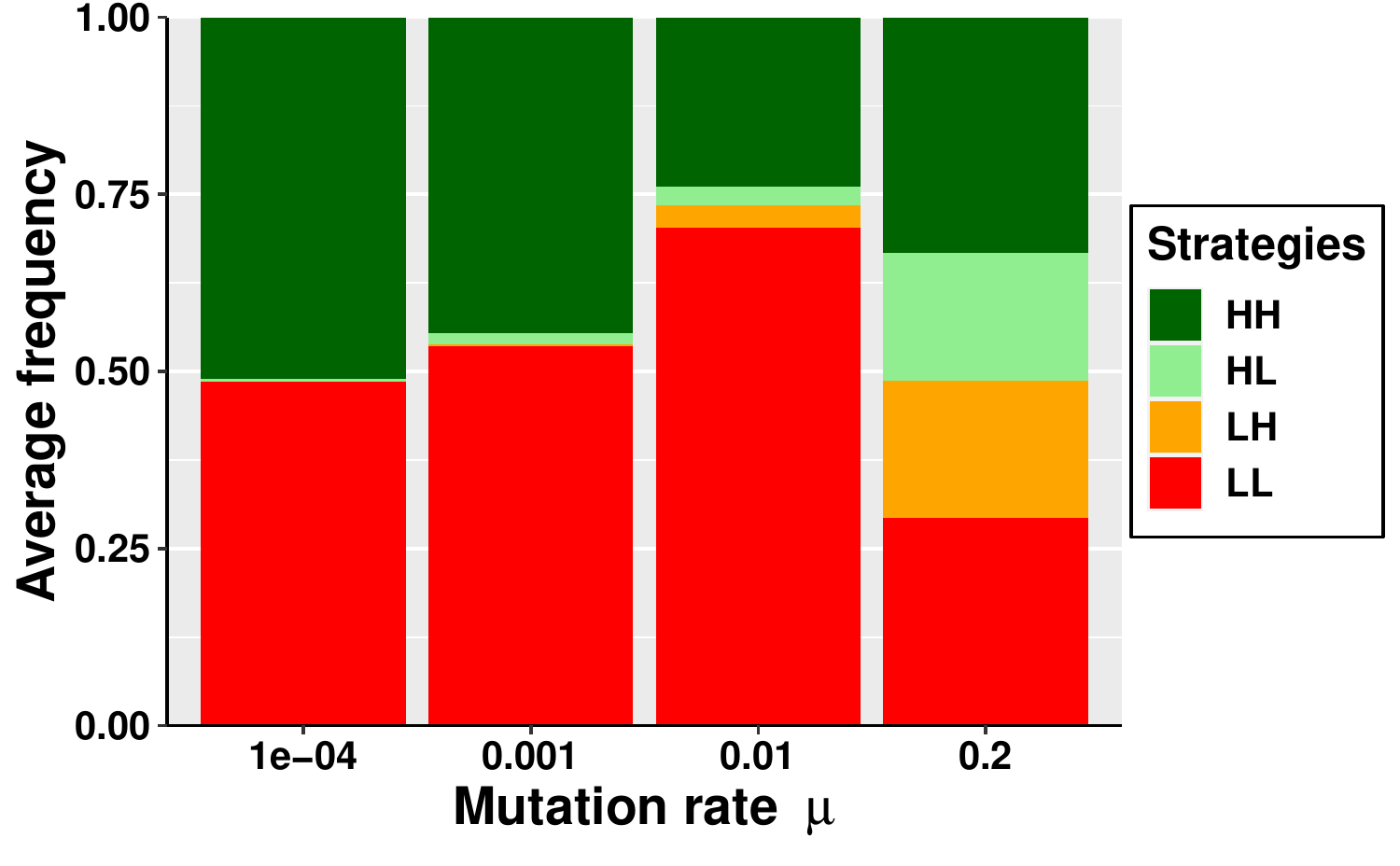}
    \caption{Average frequencies of the four  strategies HH, HL, LH and LL as a function of  mutation rate $\mu$ in absence of interference.}
    \label{fig:baseline_mut}
\end{figure}

\begin{figure}[H]
    \centering
    \includegraphics[width=0.9\linewidth]{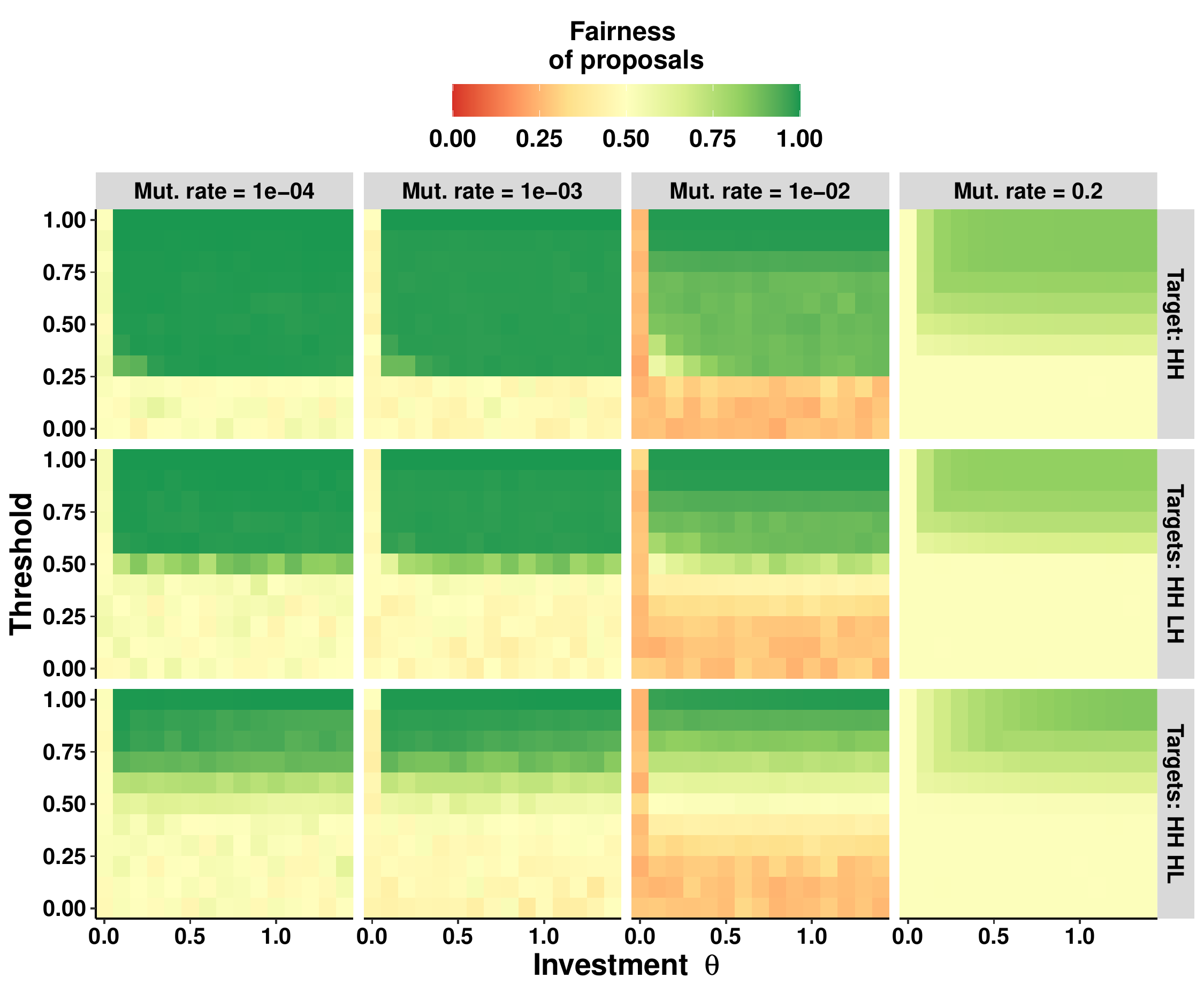}
    \caption{Average fairness as a function of the individual endowment $\theta$, the threshold $p_f$ and the mutation rate $\mu$ (population-based, stochastic update). Each row represents a different targeting scheme.} 
    \label{fig:pop_freq_cost_different_muts}
\end{figure}

\begin{figure}[H]
    \centering
    \includegraphics[width=0.9\linewidth]{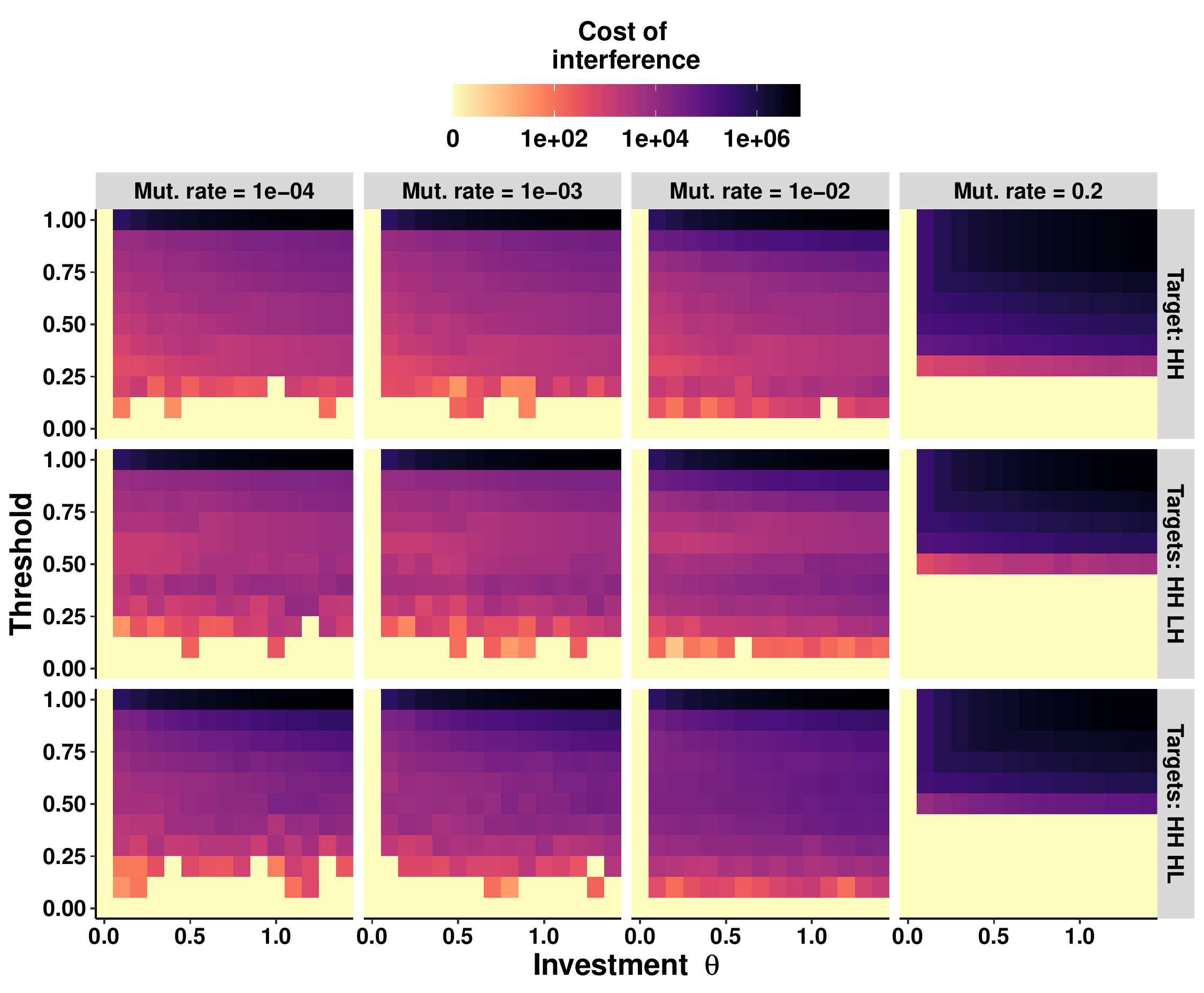}
    \caption{Average cost of interference as a function of the individual endowment $\theta$, the threshold $p_f$ and the mutation rate $\mu$ (population-based, stochastic update). Each row represents a different targeting scheme. The cost of interference is shown on a logarithmic scale.} 
    \label{fig:pop_freq_different_muts}
\end{figure}

\begin{figure}[H]
    \centering
    \includegraphics[width=0.9\linewidth]{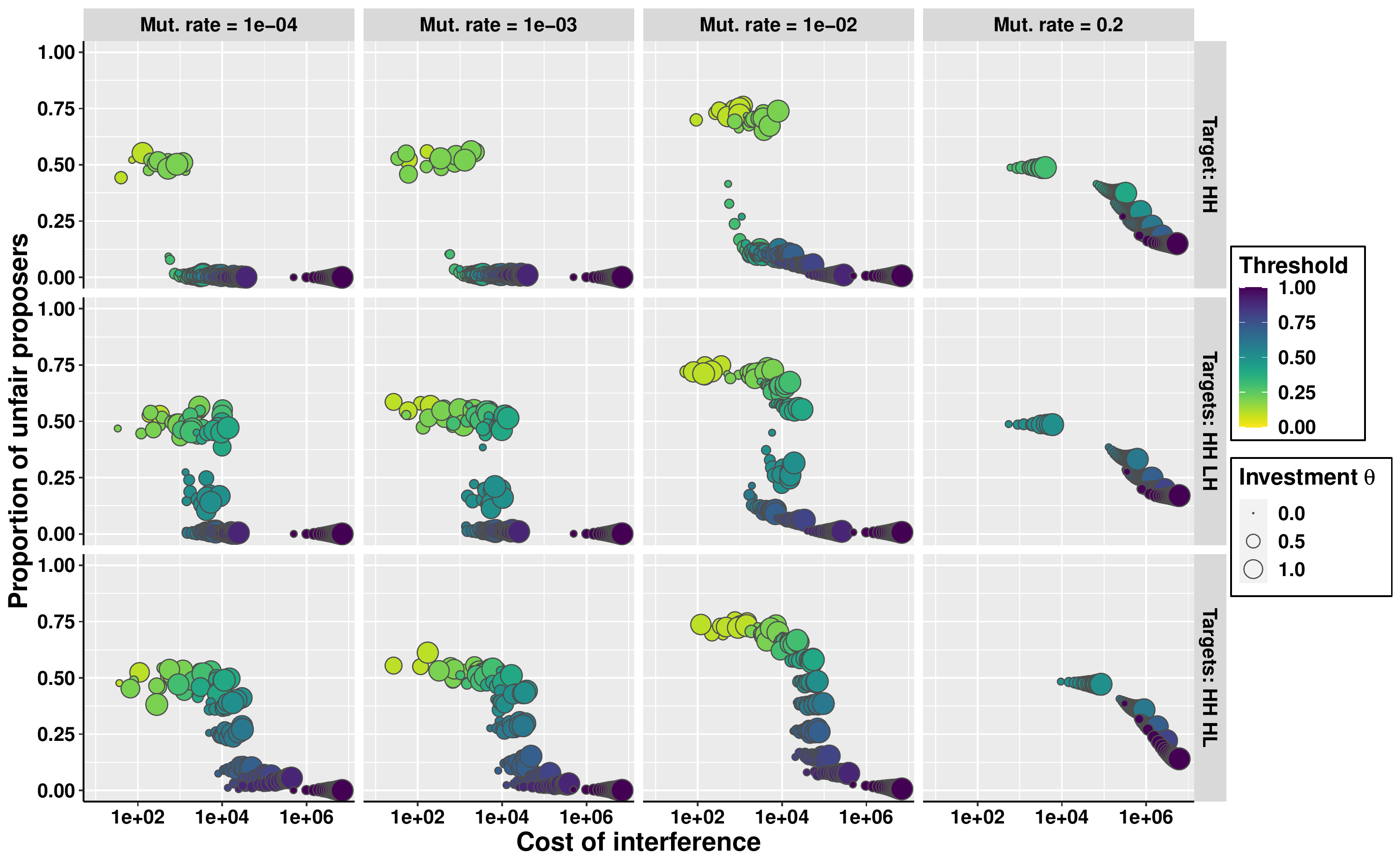}
    \caption{Proportion of unfair proposers as a function of average cost of interference for different targeting scheme and mutation rate $\mu$ (population-based, stochastic update). The size and colour of the circles correspond to investment amount and threshold of investment, respectively. We note that the most desirable outcomes are closest to the origin.}
    %I put unfair so we have a classic pareto front where the desirable outcome is on the bottom left.
    \label{fig:pop_pareto_different_muts}
\end{figure}

\begin{figure}[H]
    \centering
    \includegraphics[width=0.9\linewidth]{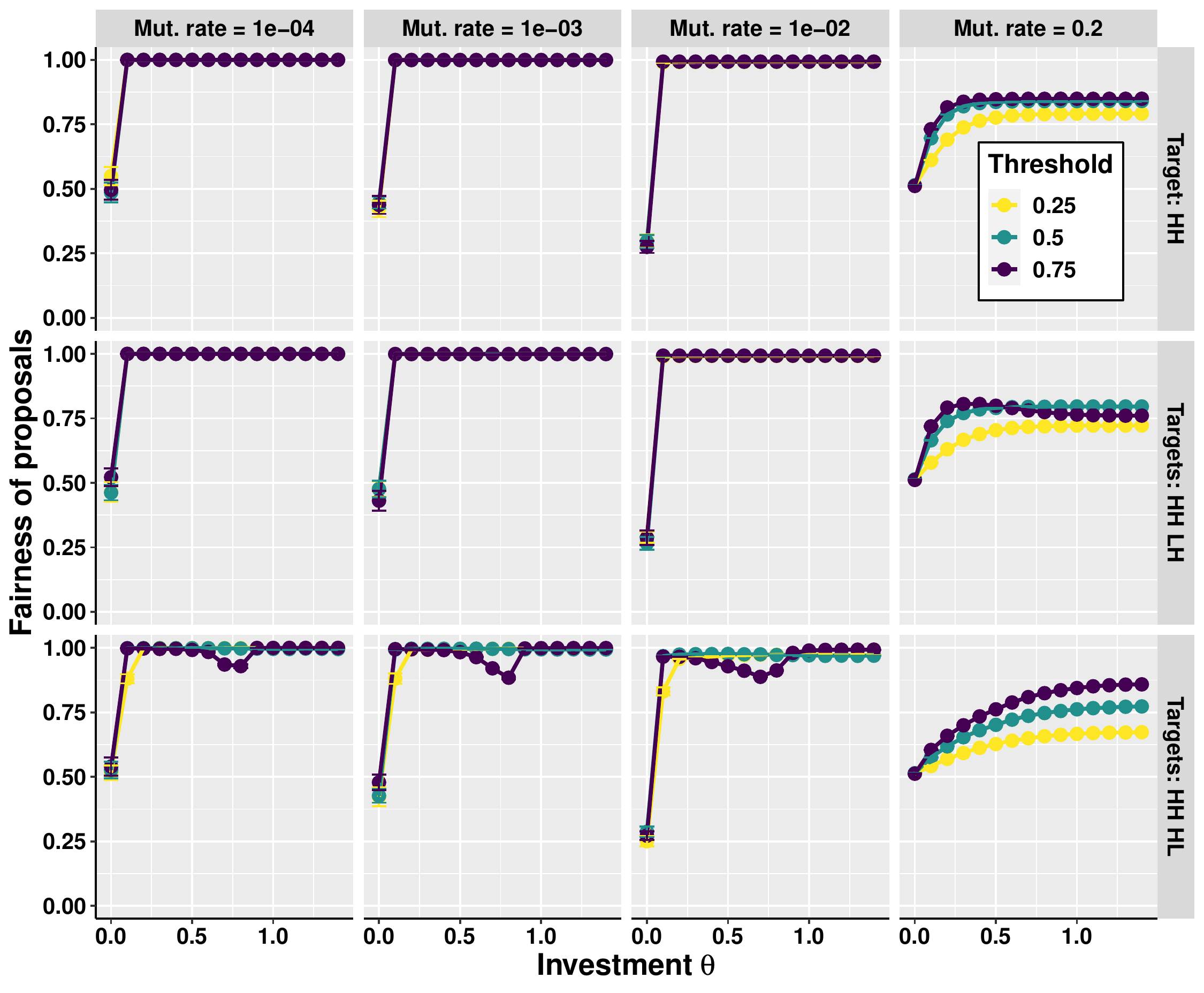}
    \caption{Average fairness measured by the sum of frequencies of HH and HL as a function of the individual endowment $\theta$, the threshold $p_f$ and the mutation rate $\mu$ (neighbourhood-based, stochastic update). Each row represents a different targeting scheme.} 
    \label{fig:neb_freq_different_muts}
\end{figure}

\begin{figure}[H]
    \centering
    \includegraphics[width=0.9\linewidth]{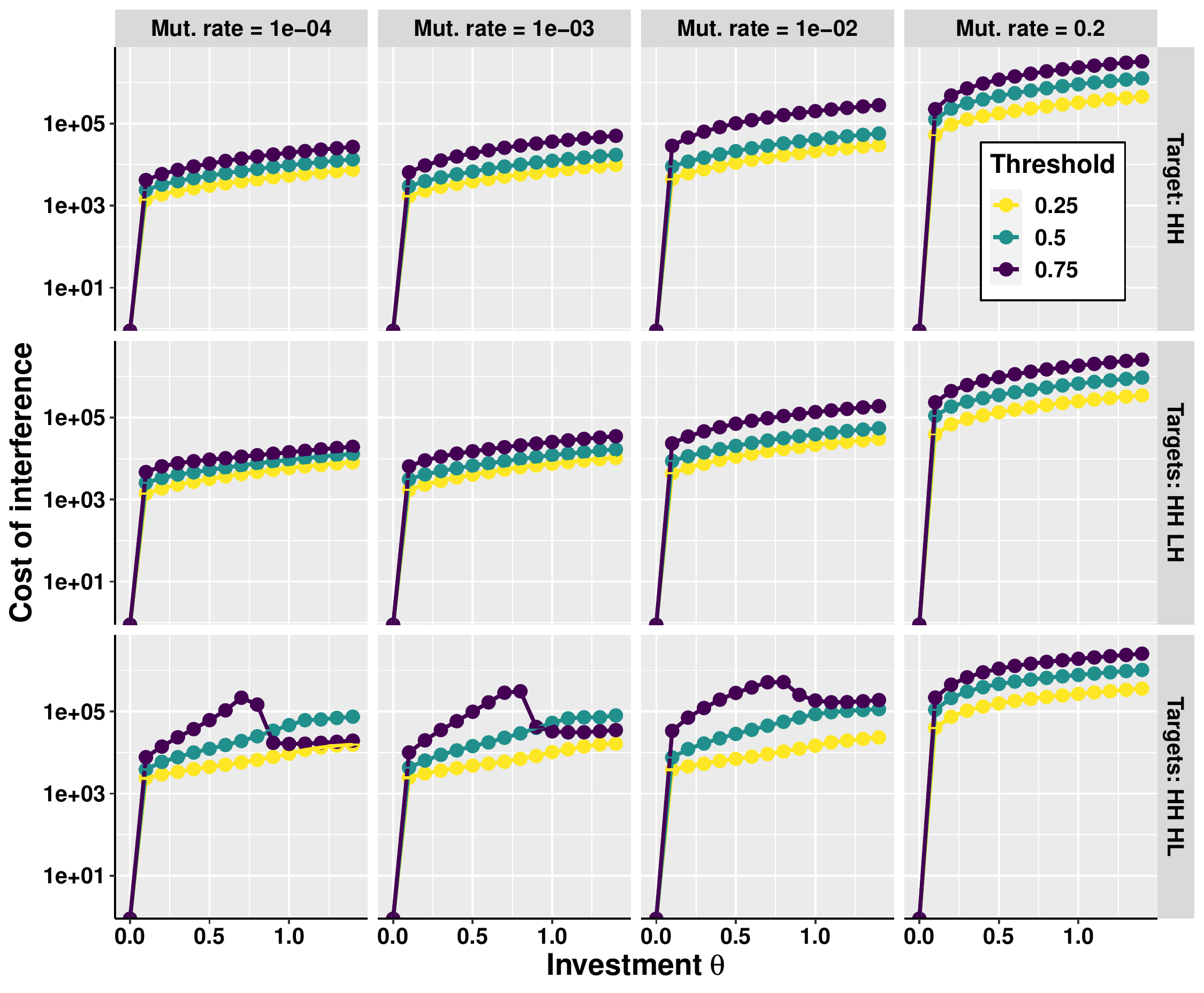}
    \caption{Average cost of interference as a function of the individual endowment $\theta$, the threshold $p_f$ and the mutation rate $\mu$ (neighbourhood-based, stochastic update). Each row represents a different targeting scheme. The cost of interference is on a logarithmic scale for clarity.} 
    \label{fig:neb_freq_cost_different_muts}
\end{figure}

\begin{figure}[H]
    \centering
    \includegraphics[width=0.9\linewidth]{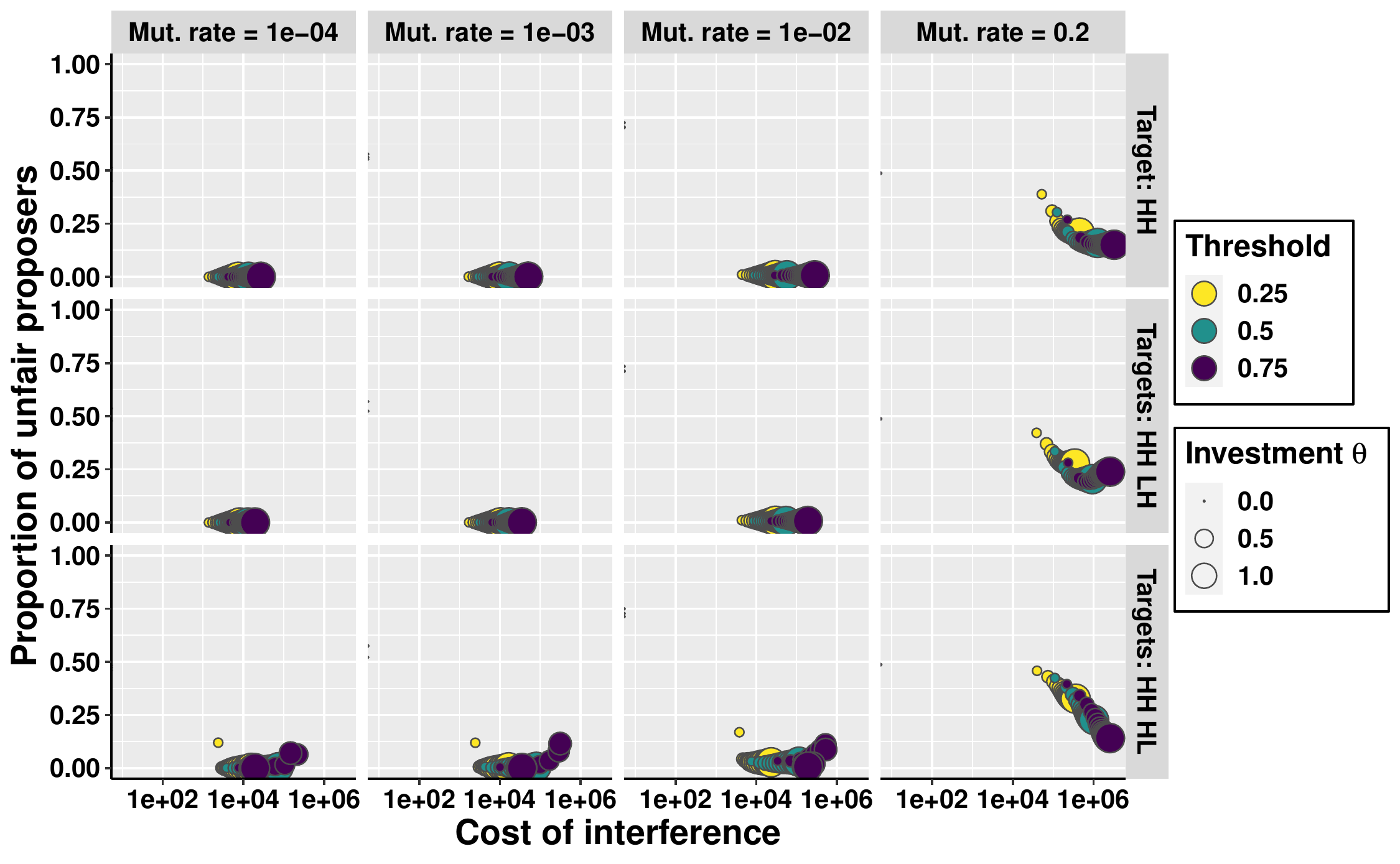}
    \caption{Proportion of unfair proposers as a function of average cost of interference for different targeting scheme and mutation rate $\mu$ (neighbourhood-based, stochastic update). The size and colour of the circles correspond to investment amount and threshold of investment, respectively. We note that the most desirable outcomes are closest to the origin.}
    \label{fig:neb_pareto_different_muts}
\end{figure}

\begin{figure}[H]
    \centering
    \includegraphics[width=0.9\linewidth]{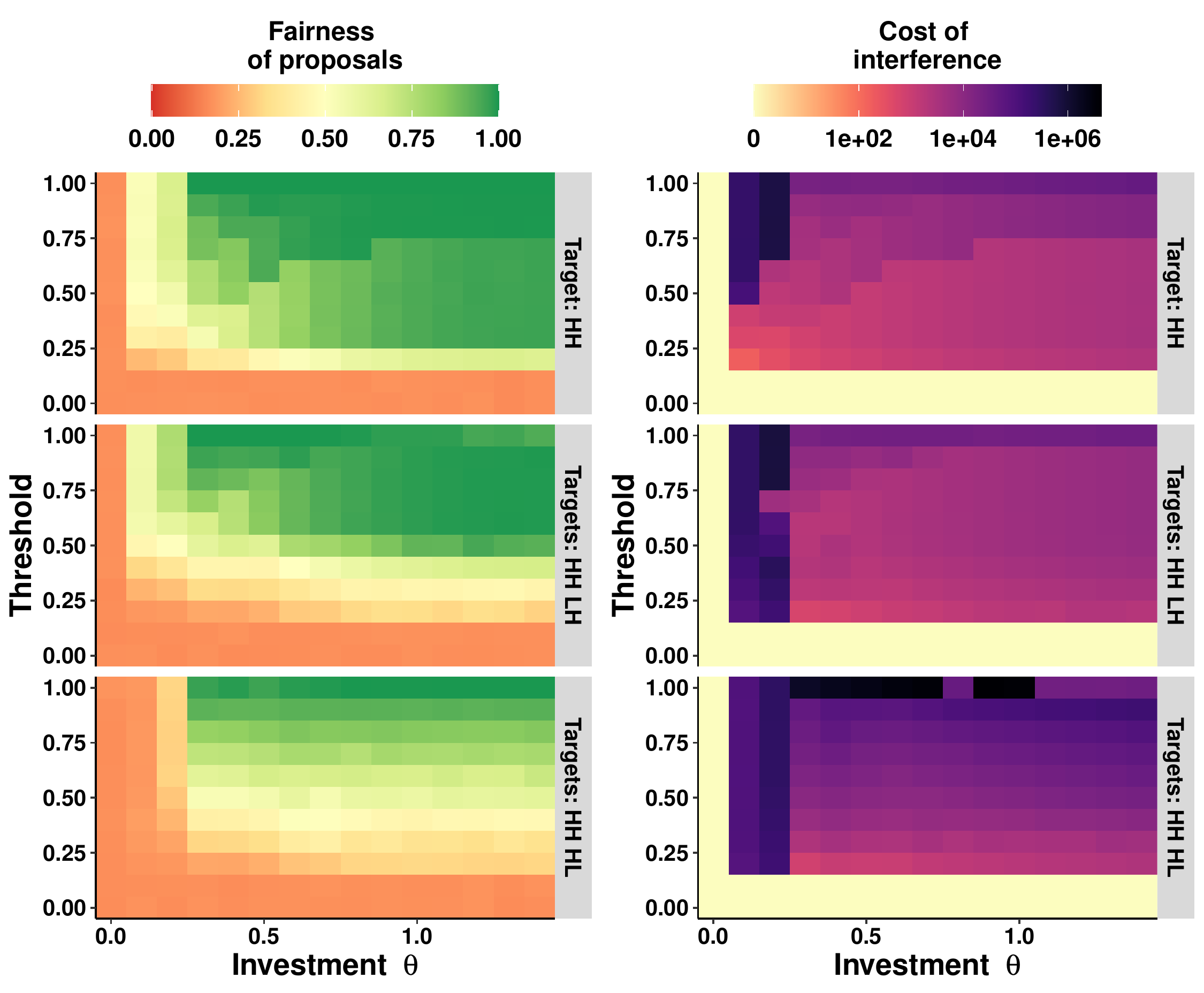}
    \caption{Average fairness (left) and average cost of interference (right) as a function of the individual endowment $\theta$ and the threshold $p_f$ (population-based, $\mu = 0.01$, deterministic update). Each row represents a different targeting scheme. The cost of interference is shown on a logarithmic scale.} 
    \label{fig:popbased_freq_cost_deterministic}
\end{figure}

\end{document}